\documentclass[journal]{IEEEtran}

\usepackage{mathrsfs}

\usepackage{graphicx}
\usepackage[fleqn]{amsmath}
\usepackage{newtxtext}
\usepackage[varg]{newtxmath}

\usepackage{comment}
\usepackage{url}
\usepackage{cite}
\usepackage{amssymb,amsfonts}
\usepackage{textcomp}
\usepackage{xcolor}
\usepackage{color}

\newcommand{\beq}{\begin{equation}}
\newcommand{\eeq}{\end{equation}}

\newcommand{\red}[1]{\textcolor{black}{#1}}

\usepackage{bm}
\usepackage{subfigure}
\usepackage[fleqn]{amsmath}
\newcommand{\argmin}{\mathop{\rm arg~min}\limits}

\begin{document}

\title{
Device-free Indoor WLAN Localization with Distributed Antenna Placement Optimization and Spatially Localized Regression
}

\author{Osamu~Muta,~\IEEEmembership{Member,~IEEE,}
        Kazuki~Noguchi,~\IEEEmembership{Student member,~IEEE,}
        Junsuke~Izumi,~\IEEEmembership{Non member,}
        Shunsuke~Shimizu,~\IEEEmembership{Student member,~IEEE,}
        Tomoki~Murakami,~\IEEEmembership{Member,~IEEE,}
        and~Shinya~Otsuki,~\IEEEmembership{Member,~IEEE}
        
\thanks{Dr. Muta is with the Center for Japan--Egypt Cooperation in Science and Technology, Kyushu University, Fukuoka, Japan.}
\thanks{
Mr. Noguchi, Mr. Izumi, and Mr. Shimizu are with the Graduate School of Information Science and Electrical Engineering, Kyushu University, Japan.
}
\thanks{Dr. Murakami and Dr. Otsuki are with NTT Corporation.}
\thanks{This work has been submitted to the IEEE for possible publication. Copyright may be transferred without notice, after which this version may no longer be accessible.}
}


\maketitle

\begin{abstract}
Wireless sensing is a promising technology for future wireless communication networks to realize various application services. 
Wireless local area network (WLAN)-based localization approaches using channel state information (CSI) have been investigated intensively. 
Further improvements in detection performance will depend on selecting appropriate feature information and determining the placements of distributed antenna elements.
This paper presents a proposal of an enhanced device-free WLAN-based localization scheme with beam-tracing-based antenna placement optimization and spatially localized regression, where beam-forming weights (BFWs) are used as feature information for training machine-learning (ML)-based models localized to partitioned areas. 
By this scheme, the antenna placement at the access point (AP) is determined by solving a combinational optimization problem with beam-tracing between AP and station (STA) 
without knowing the CSI. 
Additionally, we propose the use of localized regression to improve localization accuracy with low complexity, where classification and regression-based ML models are used for coarse and precise estimations of the target position.  
We evaluate the proposed scheme effects on localization performance in an indoor environment. 
Experiment results demonstrate that the proposed antenna placement and localized regression scheme improve the localization accuracy while reducing the required complexity for both off-line training and on-line localization relative to other reference schemes. 
\end{abstract}

\begin{IEEEkeywords}
Channel state information, device-free localization, distributed antenna, wireless local area network
\end{IEEEkeywords}

\IEEEpeerreviewmaketitle

\section{Introduction}
\IEEEPARstart{W}{ireless} sensing technology using radio signals has attracted attention as a key technology enabling the various use cases anticipated to arise during the beyond-5G and 6G era~\cite{wp_docomo, wp_kddi, ieee_jmw}. 
Such technologies are particularly expected to create new value for wireless communications by enabling wireless sensing using radio signals over wireless networks~\cite{vt_mag, com_mag, AI}.
%
%
%
Wireless sensing technology characterizes the state and behavior of a target object as fluctuations in the radio propagation environment. Particularly, many techniques have been reported using channel state information (CSI), representing the radio propagation characteristics between transmitters and receivers~\cite{wlan_sensing}. 
In addition, when combined with multi-input multi-output orthogonal frequency division multiplexing (MIMO-OFDM) transmission as multi-carrier modulation and spatial multiplexing, a large amount of CSI is expected to be acquired from wireless environments, thereby further improving wireless sensing performance such as object detection and localization accuracy.  

Various wireless sensing approaches using existing wireless interfaces have been investigated in the literature, including radio-frequency identification (RFID) sensors\cite{RFID}, ZigBee~\cite{ZigBee}, Bluetooth~\cite{Bluetooth, survey2}, and wireless local area networks (WLANs) and related studies~\cite{
survey, ref2, rss3, rss4, 
ref1, ref4, E-eyes, R-ttwd, ref3, regression1, regression2, antenna_position1, antenna_position2, antenna_position3, antenna_position4, antenna_position5, antenna_position6, antenna_position7, Murakami, Takahashi, Ishida, takata, keiretsu}. 
To be more specific, the WLAN-based approach is a promising technique that allows existing access points (APs) to be used for sensing purposes.
Broadly speaking, WLAN-based sensing techniques can be categorized into device-based and device-free approaches, depending on whether the target has a wireless device. 
%
In particular, CSI-based device-free sensing approaches (i.e., cases in which the target has no wireless device) have been studied recently, where measured CSI samples in WLANs such as IEEE802.11ac are used for wireless sensing. 
One difficulty associated with these device-free approaches is how to collect a large amount of CSI from an environment. 
To this end, a WLAN-based device-free CSI acquisition scheme was proposed in an earlier work~\cite{Murakami, Takahashi, Ishida}. For CSI monitoring, these studies used a CSI feedback mechanism for closed-loop multi-user (MU) MIMO beamforming. 
More specifically, in WLAN systems such as IEEE802.11ac, beamforming weights (BFWs) are fed back from the STA to the AP. This scheme uses a commodity WLAN interface to capture the feedback frames carrying BFWs and to extract them to train a machine learning (ML) model for wireless sensing applications such as object detection and localization~\cite{Murakami, Takahashi}.
Although this method collects a large amount of CSI effectively without explicit measurements, further investigation of the system design must be carried out to extract the potential sensing performance of WLAN systems with multiple antennas.
More concretely, the achievable detection accuracy is affected strongly by the antenna placements and their nearby radio propagation circumstances~\cite{takata}. 
However, effective antenna placement methods for device-free approaches are not well established. 
In addition, to realize WLAN-based localization with limited hardware and power resources, such as a mobile WLAN device, further investigation and analysis are necessary to develop an energy-efficient lightweight sensing solution using a small dataset.

This paper presents an effective design for device-free WLAN-based localization using beam-tracing-based antenna placement optimization and spatially localized regression. 
The original contributions of this paper are three-fold.
\begin{itemize}
\item
We propose a deterministic antenna placement optimization method without CSI. 
In this scheme, the appropriate antenna element placement at the access point (AP) is found by solving a combinational optimization problem using approximated beam-tracing information between AP and station (STA) 
without knowing the actual CSI\footnote{In other words, to determine the optimal antenna placement using CSI, it is necessary to measure CSI and train the ML models individually for all possible combinations of antenna positions, which is not a realistic approach. }, where 
the best candidate antenna placement is determined to maximize the optimization metric representing how much 
multipath-rich channel condition is present in the observation area. 
Therefore, unlike the conventional scheme~\cite{Murakami}, more accurate localization performance can be achieved by ascertaining the proper AP and STA antenna positions. 
\item
Second, we propose the use of localized regression to improve localization performance with low complexity, where a trained classification model detects the area in which a target exists and the target position is estimated more precisely using a localized regression ML model that is trained locally by CSI detected in the target area. 
Based on the property by which a simple (i.e., less complex) regression model is applicable when the localization area is small, the proposed scheme using a localized regression model is expected to improve the localization performance while reducing the required complexity. 
\item
To demonstrate the effectiveness of the proposed approaches, we conducted evaluations by experimentation using an IEEE802.11ac-based WLAN system with multiple distributed antennas in an indoor scenario, where our designed algorithms are implemented on an off-the-shelf wireless device. 
Based on the experimentally obtained results, the proposed scheme has the potential to improve localization accuracy considerably in terms of detection probability and error distance. 
\end{itemize}

\noindent{\bf Notation}:
Notation such as vectors, matrices, and variables is presented in 
Table~\ref{notation}. 

\begin{table}[t]
	\begin{center}
		\caption{List of notations}
		\label{notation}
		\begin{tabular}{ | c | l |}  \hline\hline
		{Notations and descriptions} & {Explanations} \\ \hline\hline
        {Vectors} & {Lower case letter in bold typeface} \\ \hline
        {Matrices} & {Upper case letters in bold typeface} \\ \hline
        {$\mathbf{X}^T$ } & {Transpose of a matrix $\mathbf{X}$} \\ \hline
         {$\mathbf{X}^H$ } & {Hermitian transpose of a matrix $\mathbf{X}$} \\ \hline
         {$\mathbb{R}^{a\times b}$} & {Real matrix fields of dimension $a \times b$} \\ \hline
         {$\mathbb{C}^{a\times b}$} & {Complex matrix fields of dimension $a \times b$}  \\ \hline
         {$\mathbb{Z}$} & {Set of integers} \\ \hline
         {$\min(a, b)$} & {A function that selects a smaller value} \\
         		  {}&{(either $a$ or $b$)}\\ \hline
 	{$N$} & {\red{Number of antennas at each station (STA)}} \\ \hline
		 {$M$} & {Number of antennas at APs} \\ \hline
		 {$S$} & {Number of streams} \\ \hline
		 {$K$} & {Number of subcarriers per symbol} \\ \hline
		 {$R$} & {Number of areas (labels)} \\ \hline
		 {$U$} & {Spatially concatenated CSI length} \\ \hline
		 {$d$} & {Antenna spacing at AP} \\ \hline
		 {${\bf H}_k \in \mathbb{C}^{N \times M}$} & {MU-MIMO channel matrix }\\
		 {}&{at the $k$-th subcarrier}\\ \hline
		 {$\mathbf{h}_{k,n} \in \mathbb{C}^{M \times 1}$} & {Channel vector at the $k$-th subcarrier}\\
		  {}&{and the $n$-th STA}\\ \hline
		 {$\mathbf{v}_{k,n}\in \mathbb{C}^{M \times 1}$} & {Beam-forming vector at the $k$-th subcarrier}\\
		 {}&{and the $n$-th STA} \\ \hline
		 {$\mathbf{0}^T_M \in \mathbb{C}^{M \times 1}$ } & {The zero vector with length of $M$}\\ \hline
         {$E[\cdot]$}&{Ensemble averaging} \\ \hline
		 {$_aC_b$} & {The number of combinations to select $b$ elements } \\ 
          {}&{out of $a$ elements}\\  \hline \hline
		\end{tabular}
	\end{center} 
\end{table}

\section{Related Work} 

Various WLAN and related approaches for wireless sensing have been presented in the literature~\cite{rss3, rss4, 
ref1, ref4, E-eyes, R-ttwd, ref3, regression1, regression2, antenna_position1, antenna_position2, antenna_position3, antenna_position4, antenna_position5, antenna_position6, antenna_position7, Murakami, Takahashi, Ishida, takata, keiretsu}. 
Radio signal strength (RSS) based sensing techniques have been widely investigated because RSS data acquisition is more accessible than other approaches. 
For one earlier study \cite{rss3}, a clustering algorithm was proposed to remove noise samples for RSS fingerprint-based WLAN indoor positioning, where a DBSCAN-based clustering scheme was introduced. 
Another earlier study~\cite{rss4} examined a proposed alternative fingerprint-based localization method, the data-rate-based fingerprint framework, in which the transmission power and the achieved data rate are used as fingerprint information. This method achieves localization accuracy that is comparable to that obtained using RSS-based methods. 
%
The authors in~\cite{ref4} propose a distributed massive MIMO-based localization scheme using RSS data clustering to reduce the required complexity, where large RSS measurements are used to estimate a user's location.
As discussed in the existing studies described above, RSS and other related information are easily exploited for sensing purposes in most off-the-shelf WLAN devices. 
However, RSS is the average power measurement over the signal bandwidth. It is inadequate for obtaining more accurate sensing results. 
CSI is more effective for improving measurement accuracy because it shows the channel characteristics' impulse response or frequency response. 
Various studies have used methods based on a device-free concept, where a target object has no wireless device. 
One report presents ~\cite{E-eyes} a device-free WLAN-based location-based activity identification scheme. To identify various activities, such as in-place activities and walking movements in a house, this scheme exploits CSI in OFDM systems. 
Another report ~\cite{R-ttwd} presents a through-the-wall human detection system using off-the-shelf WLAN devices, where principal component analysis-based filtering is applied to extract meaningful features from the CSI. More specifically, the correlated subcarriers are selected to extract features for robust human detection, whereas a convolutional neural network-based approach has been applied for CSI-based positioning in another study~\cite{ref3}. 
Device-free regression-based localization schemes are presented in some reports~\cite{regression1, regression2}, where support vector regression is used to localize the target. 
One report ~\cite{regression2} describes that solving a device-free localization problem with regression models is more effective than when solving a classification problem. 
However, the required complexity for constructing an accurate regression model might be much higher than the case using the classification model.  
The approaches described in those reports rely on the available CSI as feature information for wireless sensing purposes. Consequently, the achievable sensing accuracy depends on the antenna placements and their surrounding radio propagation environments. 
Although antenna placement problems have been considered for coverage prediction and optimization ~\cite{antenna_position1}, how to determine antenna placements that improve sensing accuracy in device-free WLAN-based systems has not been investigated sufficiently. 
In some studies~\cite{antenna_position2, antenna_position3}, a high-frequency antenna array has been used to locate partial discharge sources.
As reported~\cite{antenna_position3} for one other method, the antenna array placement is optimized heuristically to locate the signal source. 
Nevertheless, these techniques are not directly applicable to device-free scenarios because no signal source or wireless device is present at the target. 
Also, AP selection algorithms have been investigated for WLAN-based localization~\cite{antenna_position4, antenna_position5, antenna_position6, antenna_position7}. 
Offline and online AP selection strategies are presented in~~\cite{antenna_position4}, where the best AP combination among all possible AP sets is selected. 
Authors in~\cite{antenna_position5, antenna_position6} proposed a mixed offline and online AP selection for improving the noise tolerance of an ML model. 
Authors in~\cite{antenna_position7} proposed an access point optimization approach to deploy the given number of WLAN APs in an indoor environment, where simulation-based analysis is given for performance evaluation. 
However, the schemes described above rely on RSS measurement at the target. Therefore, the feature information must be measured in advance. In addition, AP selection for device-free-based localization is not discussed. 
Unlike the schemes above, this paper is intended to develop an effective antenna placement scheme without knowing feature information, i.e., without measuring CSI.  

Another difficulty is how to acquire a large amount of CSI from the environment. 
As realistic solutions, earlier reports ~\cite{Murakami, Takahashi, Ishida, takata, keiretsu} have proposed effective CSI acquisition schemes where a CSI feedback mechanism for IEEE802.11ac-based WLANs is used to collect CSI from  all nearby devices as feature information for sensing purposes.
As a result, the sensing area can be expanded easily without installing additional sensing stations (STAs).
As described in one earlier report ~\cite{Takahashi}, the collected CSI samples are used for WLAN-based human detection with a deep neural network and numerous CSI samples. Unlike deep learning-based methods trained on large datasets, the method described herein is a lightweight approach using a small dataset.

Unlike those earlier studies, we aim for this study to design an effective distributed antenna placement method that determines antenna placements deterministically without knowing the CSI dataset. 
In addition, to achieve further performance improvement, we propose a localized regression-based method that works with a small dataset. After coarsely detecting the local area where the target is located, a target position is estimated using a locally optimized regression model to the detected target area. 
Based on experimentally obtained results in an indoor distributed MIMO scenario, we clarify the localization performance of the scheme proposed above. 


\begin{figure}[t] 
\begin{center}
\includegraphics[width = 8.5cm]{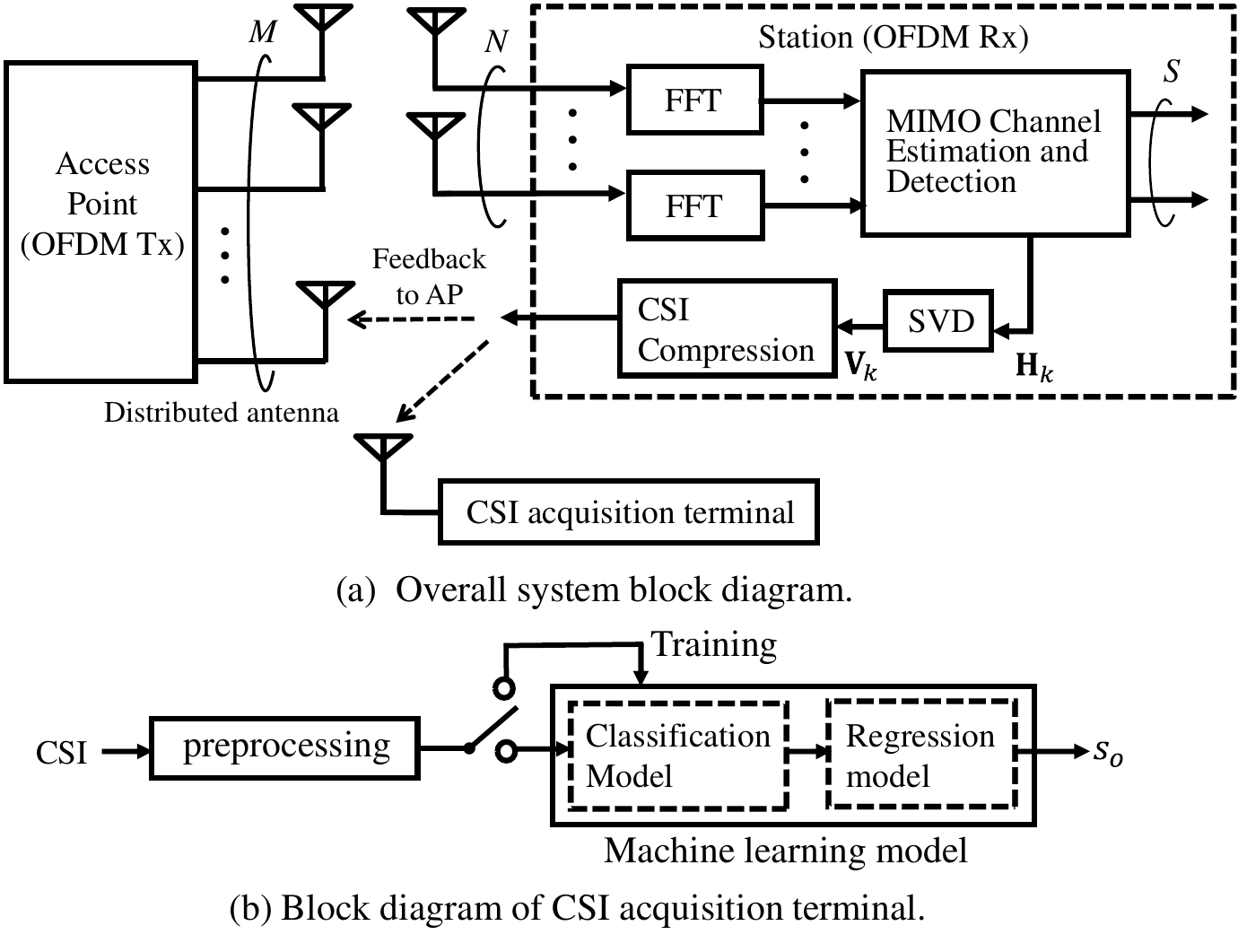} 
\caption{Block diagram of a WLAN-based localization system in which AP with $M$ antennas serves STA with $N$ antennas. The CSI-acquisition terminal extracts BFWs from captured feedback frames and applies them to an ML model consisting of classification and regression models, respectively, for area detection and area-wise localization.}
\label{Block diagram}
\end{center}
\end{figure}

\section{System Descriptions\label{System_Model21}}
A comprehensive block diagram of the device-free WLAN-based localization system using the proposed scheme is portrayed in Fig.~\ref{Block diagram}, where an IEEE802.11ac-based AP with $M$ antenna serves a user device (STA) with $N$ antennas. Here, the number of streams is $S= \mbox{min}(M, N)$.
In device-free localization systems, the target has no wireless device. 
Here, a CSI acquisition terminal captures feedback frames from STA to AP and then extracts the BFWs, which are used to train the regression and classification ML models for coarse and fine target location estimation. 
Details of the ML models used for this study are explained hereinafter. 

Assuming that positions of $M$ AP antennas and the STA are determined properly in advance, the
channel matrix of the path between the AP and the STA at the $k$-th subcarrier is represented by  
${\bf H}_k =\left[\mathbf{h}_{k,1}, \cdots, \mathbf{h}_{k,N}\right]\in \mathbb{C}^{N \times M}$, 
$k=1,\cdots,K$, where $K$ denotes the number of subcarriers per OFDM symbol.
Here, $\mathbf{h}_{k,n}=[h_{k,n,1}, \cdots, h_{k,n,M}]^T \in \mathbb{C}^{M \times 1}$, where 
$h_{k,n,m}$ is the channel coefficient between the $m$-th transmit antenna and the $n$-th receive antenna at the $k$-th subcarrier.
Consequently, the concatenated channel matrix over subcarriers in the frequency domain can be expressed as ${\bf H}=[{\bf H}_1, \cdots, {\bf H}_k, \cdots, {\bf H}_K]$.

The CSI per subcarrier is estimated at the receiver side (STA). 
After OFDM demodulation with FFT processing, the channel matrix at the $k$-th subcarrier is estimated as $\hat{{\bf H}}_k \in \mathbb{C}^{N \times M}$. 
The right singular matrices, ${\bf V}_k \in \mathbb{C}^{M \times M}$, are obtained by singular value decomposition (SVD) of $\hat{{\bf H}}_k$, i.e., the channel matrix is decomposed into $\hat{{\bf H}}_k = {\bf U}_k\Sigma_k {\bf V}_k^H $, where $\Sigma_k \in \mathbb{C}^{N \times M}$ denotes a diagonal matrix in which the diagonal element is a singular value of the channel.
If $M>N$, then the right singular matrix is reduced to ${\bf V}_k =\left[ \mathbf{v}_{k,1}, \cdots, \mathbf{v}_{k,N}, \mathbf{0}^T_M, \cdots, \mathbf{0}^T_M \right] \in \mathbb{C}^{M \times M}$. 
The first $N$ column vectors $\mathbf{v}_{k,n}$ are used as downlink BWFs for $N$ user devices.
Here, $\mathbf{v}_{k,n}=[v_{k,n,1}, \cdots, v_{k,n,M}]^T \in \mathbb{C}^{M \times 1}$ and $\mathbf{0}^T_M = [0, \cdots, 0]^T \in \mathbb{C}^{M \times 1}$ denotes the zero vector of length $M$.
Based on the IEEE802.11ac standardized schemes, the right-singular matrix ${\bf V}_k$ is compressed by application of Givens rotation as a linear transform and quantized~\cite{IEEE}~\footnote{The right singular matrix ${\bf V}_k$ is converted into a quantized version of an angle information sequence. The received angle information sequence is converted to the right-singular matrix at the CSI acquisition terminal. Details are presented in Appendix A of Ref.~\cite{keiretsu}.}.
The generated CSI, i.e., the quantized and compressed version of the right-singular matrix, is fed back from the STA to the AP.  

\begin{figure}[t] 
\begin{center}
\includegraphics[width = 8.5cm]{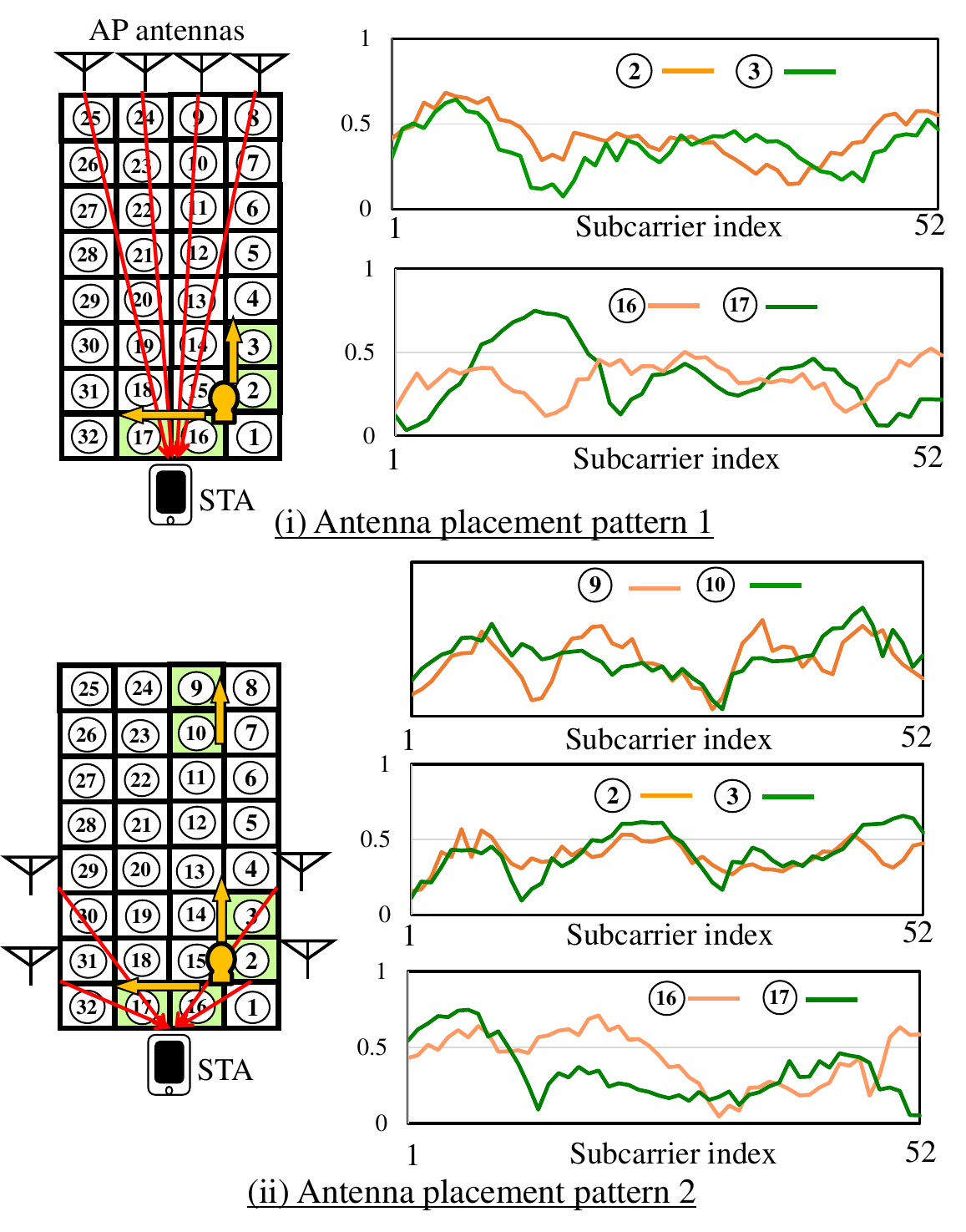} 
\caption{
\red{Examples of measured CSI when the target is located at different positions in two antenna placement cases (i) and (ii), respectively.}
}
\label{Measured CSI}
\end{center}
\end{figure}

Fig.~\ref{Block diagram}(b) depicts a block diagram of the CSI acquisition terminal to capture the feedback frames and then to reconstruct the right-singular matrix (i.e., BFW) $\tilde{{\bf V}}_k$ as feature information. 
At the preprocessing block, consecutively received BFWs are concatenated as more effective single-feature information. 
To be more specific, let $\hat{\mathbf{V}}^{(p)}_k$ denote the BFW matrix at the $p$-th time instance on the $k$-th subcarrier. 
The BFW matrix over subcarriers is represented as $\hat{\mathbf{V}}^{(p)}=[\hat{\mathbf{V}}^{(p)}_1, \cdots, \hat{\mathbf{V}}^{(p)}_K]$. 
The concatenated BFW matrix is given as  
\begin{equation}
\hat{\mathscr{V}}^{(p)}=\left[\hat{\mathbf{V}}^{(p-(U-1))}, \cdots, \hat{\mathbf{V}}^{(p)}\right], \hspace{5pt} p>U, 
\end{equation}
where $U$ is defined as the concatenated CSI length (i.e., the number of concatenated BFWs), 
which is used as effective feature information for ML-based localization~\cite{keiretsu}.

\section{Proposed Localization Scheme\label{proposal}} 
This section presents an explanation of the concepts of the proposed localization scheme based on beam-tracing-based antenna placement optimization and spatially localized regression.

Fig.~\ref{Measured CSI} portrays two examples of antenna placements 
in an indoor environment, where four AP antennas are placed on the different sides of STA. 
The measurement region is divided into $R=32$ areas. 
Figs.~\ref{Measured CSI}(i) and ~\ref{Measured CSI}(ii) respectively show the target moving in measurement areas with different antenna placements.
%
%
In Fig.~\ref{Measured CSI}(i), when the target moves across the direct paths, i.e., from area 16 to 17, it tends to fluctuate the channel characteristics between AP and STA. 
%
By contrast, 
when the target moves from area 2 to 3, i.e., approximately along with the direct beam between AP and STA, 
the channel fluctuation weakens even when the target moves. 
By contrast, in Fig.~\ref{Measured CSI}(ii), similarly to the situation described above, the CSI fluctuates when the target moves from area 16 to 17, although it is less affected when the target is in area 2 to 3.
The upper right panel of this figure shows that CSI varies even when the target is located in areas with no direct path between AP and STA, such as areas 9 and 10.  This finding indicates that multi-paths from the walls characterize the target's behavior and consequently facilitate target detection. 
%
%
This fact implies that higher localization accuracy can be expected when antenna elements are placed to reduce the coverage areas of direct beams and when they make the environment become multi-path rich. 
The proposed scheme uses this concept to elucidate the appropriate antenna placement without knowledge of the actual CSI~\footnote{
\red{
The proposed scheme determines the AP antenna position without knowing the actual CSI between AP and STA. In other words, the proposed scheme implicitly determines the AP antenna positions based on the assumption that the transmit beamforming is used at the AP, i.e., the transmit signal beam at the AP is directed to the STA side.  
}
}. 

\begin{figure}[t]  
\begin{center}
\subfigure[With metric $S_1$]
{
\includegraphics[width = 3.0cm]{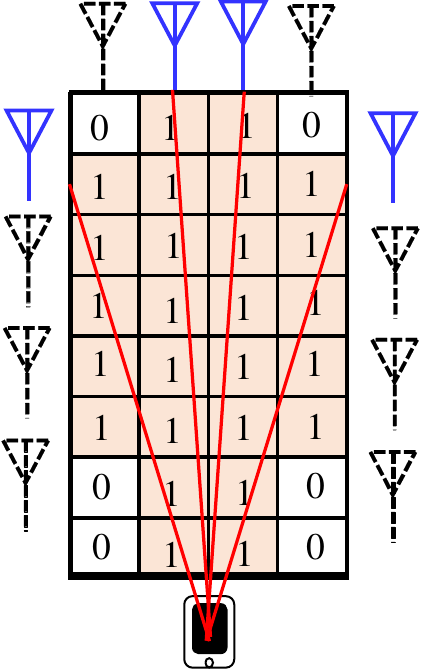}
\label{S1}
}
\hspace{10pt}
\subfigure[With metric $S_2$]{
\includegraphics[width = 3.0cm]{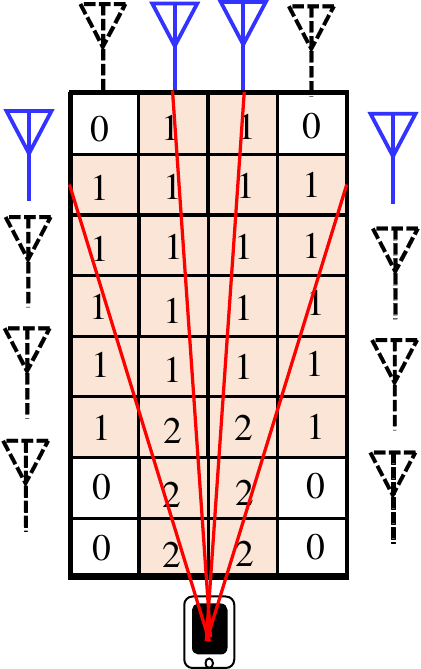}
\label{S2}
}

\subfigure[Actual room layout and reflected wave trajectory]
{
\includegraphics[width = 2.0cm]{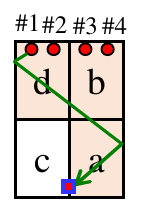}
\label{S3}
}
\hspace{10pt}
\subfigure[Representation of reflected wave trajectory using the actual room and its mirror image rooms]{
\includegraphics[width = 4.0cm]{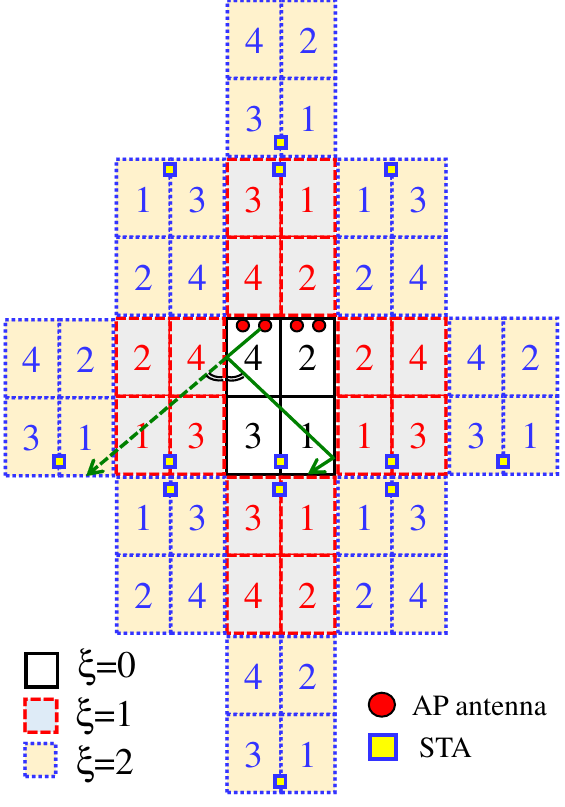}
\label{S4}
}

\caption{Illustrations of antenna position optimization based on the beam trajectory tracing between AP ($M=4$) and STA:
(a) and (b) respectively portray how to calculate two metrics $S_1$ and $S_2$. 
Also, $\xi=0$ and $\xi\ge 1$ respectively denote the actual room and its mirror image rooms.}
\label{S1andS2}

\end{center}
\end{figure}

\red{
This paper presents consideration of two evaluation metrics, $S_1$ and $S_2$, respectively, for antenna placement design. 
Instead of knowing the actual CSI, the proposed scheme uses two metrics to evaluate the antenna placement patterns. 
Metrics $S_1$ and $S_2$ are to evaluate coverage areas by direct beams and the number of beams in the observation field, respectively. 
In that sense, metric S1 is defined as the number of areas through which direct beams pass, while the second metric $S_2$ is defined as the total number of direct beams through which direct beams pass. Concrete examples to explain the difference between the two metrics are shown in Figs.~\ref{S1andS2}(a) and ~\ref{S1andS2}(b)}
Four blue antennas show selected antenna positions among the 12 candidates in these figures. Red lines show direct beams between AP antennas and STA, where the areas at which the beams pass are colored red. 
As in Fig.~\ref{S1andS2}(a), the first metric $S_1$ is the number of red areas, i.e., the number of areas through which the beam passes. 
The second metric $S_2$ in Fig.~\ref{S1andS2}(b) is the total number of beams passing over red areas. 
Metric $S_2$ can be extended to cases with reflected waves from walls and obstacles.  
Figs.~~\ref{S1andS2}(c) and \ref{S1andS2}(d) show the actual room layout and its extended layout with mirror images of the actual one, where the green line shows beam-tracing between AP antennas and STA. 
Here, $R=4$ is assumed with labels of $a$, $b$, $c$, and $d$. 
An example of a reflected wave trajectory in a room is presented in Fig.~\ref{S1andS2}(c). 
The metric $S_2$ is the difference between the number of direct beams and that of reflected beams passing through red areas.
This metric is calculable using mirror images of the actual room, as shown in Figs.~\ref{S1andS2}(d).
The red and blue boxed areas of this figure respectively present mirror images of $\xi=1$ and $2$. 
Here, $\xi$ denotes the $\xi$-th reflected wave, i.e., $\xi=1$ and $2$ respectively represent the trajectories of the first and second reflected waves.
This figure clarifies that $S_2$ is calculable in the same way as the direct waves, i.e., by adding up the number of beams passing through the colored areas with the same label ($a$, $b$, $c$, and $d$). 
For example, in Fig.~\ref{S1andS2}(d), the first and second reflected waves respectively pass over areas $d$, $b$, and $a$ in the mirrored rooms of $\xi=1$ and area $a$ in that of $\xi=2$.
As the direct wave path through area "d", $S_2=5$ is obtained by adding them up. 
The scheme above can be extended to a case with an arbitrary number of areas, e.g., $R=32$~\footnote{The number of area divisions in antenna placement determination is not necessarily the same as $R$ in the classification. 
To simplify the discussion, for these analyses we assume the same $R$ for calculating $S_1$ and $S_2$. }.

Let $\xi=0, 1, \cdots, \Xi$ denote the mirror image index corresponding to the $\xi$-th reflected wave, where $\xi=0$ stands for the direct beam, and where $\Xi$ denotes the maximum number of reflections per path between AP antenna and STA.  
Two metrics $S^{[b]}_1$ and $S^{[b]}_2$ are defined as 
%
\beq
S^{[b]}_1=
\sum_{r=1}^{R} g(c_r^{0,b})
\label{Eq2}
\eeq
\beq
S^{[b]}_2=
\sum_{r=1}^{R} c_r^{0,b}-
\sum_{\xi=1}^{\Xi}r_{\xi}\sum_{r=1}^{R} c_r^{\xi,b}, 
\label{Eq2b}
\eeq
where $b$ stands for the index of the selected antenna placement pattern, and where  
$r_{\xi}$ is an attenuation factor of reflected waves, i.e., $r_{0}=1$ and $r_{\xi}\le 1, \forall \xi$. 
In those equations, $c_r^{\xi,b}$ denotes the number of times that the $\xi$-th beam passes through area $r$ when antenna placement pattern $b$ is used, where $\xi=0$ and $\xi\ge 1$ respectively correspond to direct and $\xi$-th reflected beams. 
Also, $g(\cdot)$ is a binary decision function defined as 
\begin{equation}
g(x)= \left\{
\begin{array}{cc}
     1 &  x>0 \\
     0 &  x=0
\end{array}
\quad x\ge 0. 
\right.
\end{equation}
The first and second terms on the right-hand side of the equation (\ref{Eq2b}) respectively correspond to direct beams and reflected beams. 
\red{The second term on the right-hand side is the impact of reflected beams.}
The sign of the second term is opposite to that of the first term. Consequently, minimizing $S_2^{[b]}$ leads to reduction of the number of direct signals, but it also increases the number of reflected signals. 
\red{
We consider both $S_1^{([b])}$ and $S_2^{([b])}$, simultaneously by considering multiplication of both metrics in the formulated problem.
}
%

We formulate the antenna position determination problem as a combinational optimization problem that minimizes the 
product of $S_1$ and $S_2$. 
The formulated problem is given as  
\begin{eqnarray}
\argmin_{b} &\hspace{10pt}& S^{[b]} 
=S^{[b]}_1 S^{[b]}_2, 
\hspace{5pt}
b\in \mathscr{B}
\label{p1a}\\ 
s.t. 
&\hspace{10pt}& S_1^{[b]}>0, \label{p1b}\\
&\hspace{10pt}& M \le \mathscr{M}, \label{p1c}
\end{eqnarray}
Therein, $\mathscr{B}$ represents the set of antenna placement patterns. 
The cardinality of $\mathscr{B}$ is $|\mathscr{B}|=_\mathscr{M}C_{M}$, where $M$ ad $\mathscr{M}$ denote the number of selected AP antennas and that of the candidate positions, respectively. 
Here, the objective function is the geometric mean of two metrics: $S^{[b]}_1$ and $S^{[b]}_2$, i.e., minimizing $S^{[b]}_1$ maximizes the areas through which the main beams never pass, whereas minimizing $S^{[b]}_2$ minimizes the total number of the direct beams passing through the observation areas while maximizing the number of reflected beams. 
The geometric mean is used to incorporate both viewpoints simultaneously. 
(\ref{p1b}) represents constraints by which metric $S_1^{[b]}$ is a positive value. 
(\ref{p1c}) signifies that selected antennas $M$ never exceed the maximum value $\mathscr{M}$. 
The best antenna position to minimize (\ref{p1a}) is selected among all possible ones $\mathscr{B}$, e.g., when $\mathscr{M}=12$ and $M=4$, the number of antenna position combinations is $|\mathscr{B}|=_{12}C_{4}=495$~\footnote{\red{
Based on the assumption that the appropriate antenna pattern is determined without knowing the actual CSI in advance, it is not necessary to solve the problem in real-time. 
Although this paper uses an exhaustive search to solve (\ref{p1a}), if a more efficient solution approach is needed, existing relaxation algorithms, e.g., for channel assignment algorithms in wireless communications\cite{Elwekeil1, Elwekeil2}, may be applied. 
For further investigation with relaxation algorithms is one of our future study items.
}}. 
%

\begin{figure}[t] 
\begin{center}
\includegraphics[width = 8.5cm]{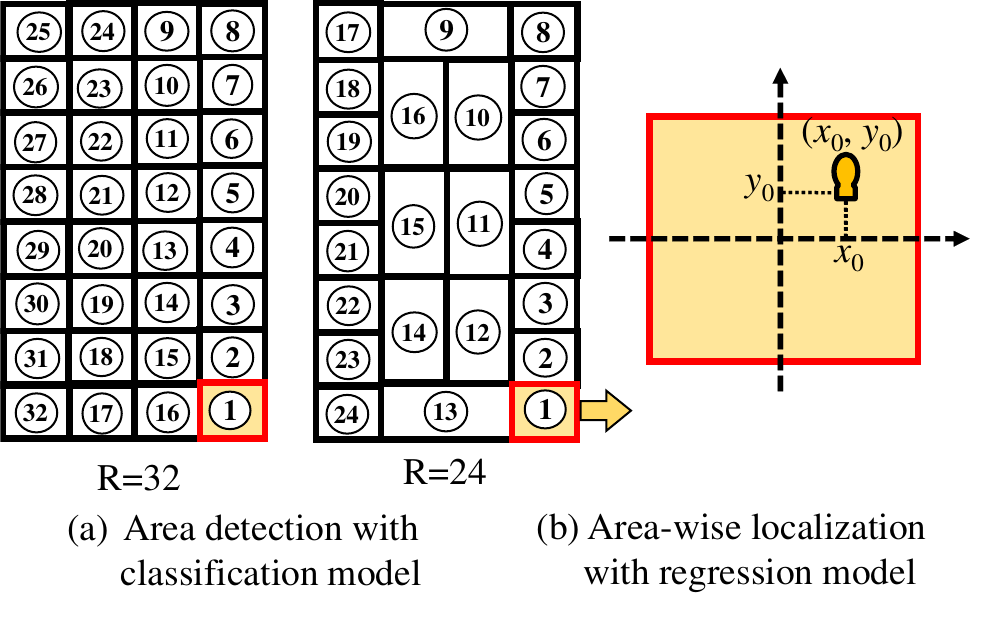} %
\caption{Illustration of target position localization in which the multi-class classification model and regression models are used respectively for area (label number) detection and area-wise localization.}
\label{Fig4}
\end{center}
\end{figure}


To improve the localization accuracy further with low complexity after the appropriate antenna position is given, we propose the use of a spatially localized regression model for more effective localization, where trained classification and regression ML models are used, respectively, for coarse and precise detection of the target position. 
To be more specific, after the area in which the target is located is detected using a trained classification model, a spatially localized area-wise regression model is used to estimate the target position in the detected area. 
Fig.~\ref{Fig4} presents the concepts of a spatially localized area-wise regression localization: a target's two-dimensional coordinates $(x_0, y_0)$ are estimated using a localized regression model, whereas the target area (\#1 in Fig.~\ref{Fig4}(a)) is detected using a classification model, where the regression model is trained individually, area-by-area. 
Let ${\bf V}_k^{(i,r)}\in \mathbb{C}^{M\times M}$ denote the $i$-th feature information at area $r$ for training both localized regression and classification models, where $k$ denotes the subcarrier index. 
The training data samples for the regression and classification models are represented, respectively, as
\begin{eqnarray}
\{({\bf V}_k^{(i,r)},(x^{(r)}_i,y^{(r)}_i))\},\ \ \ \ \ i=1,...,\mathscr{N}
\label{regress1}
\end{eqnarray}
where $(x^{(r)}_i, y^{(r)}_i)$ are the correct data of the $i$-th two-dimensional coordinate of the target's position at area $r$. 
In that expression, $\mathscr{N}$ denotes the number of training data samples per area. 
%
In the regression model, the $r$-th localized model is trained to map the feature information (i.e., ${\bf V}_k^{(i,r)}$) to the correct two-dimensional coordinate $(x^{(r)}_i, y^{(r)}_i)$ when the target is in area $r$. 
\red{To be more specific, in random forest for regression, the mean square error between prediction function $f({\bf V}_k^{(i,r)})$ for the training dataset ${\bf V}_k^{(i,r)}$ and the correct data Y is formed as mean square error of distance error~\cite{RF}, i.e.,}
$$
\red{E[((x_i^{(r)},y_i^{(r)})- f({\bf V}_k^{(i,r)}))^2]},
$$
\red{where $E[ ]$ denotes expectation operator~\footnote{
\red{
In Sect. V, we used the scikit-learn library in Python to build the trained random forest model.
}
} (For more details, see Ref.~\cite{RF}).} 
%
%
When the localization region is divided into $R$ multiple sub-regions (called "areas"), the $R$ individual localized regression models are constructed using the feature information labeled to $r=1, \cdots, R$. 
Then, the target localization is performed with the trained regression model for the corresponding area. A multi-class classification model is used to determine which localized regression model is used (e.g., in Fig.~\ref{Fig4}, the trained model for area 1 is used)~\footnote{
\red{
CSI (right-singular matrix of channel matrix H) is generated at the STA for the beamforming purpose at the AP side. The proposed scheme uses the beam-forming weights as feature information for the target localization. The difference between the behavior of the channel matrix H and that of the beamforming matrix V has been explained in our previous study in Ref.~\cite{keiretsu}.
}
}.

\section{Performance Evaluation\label{performance_evaluation}} 

\subsection{Experiment scenarios and setup}

\begin{figure}[t] 
\begin{center}
\includegraphics[width = 8.5cm]{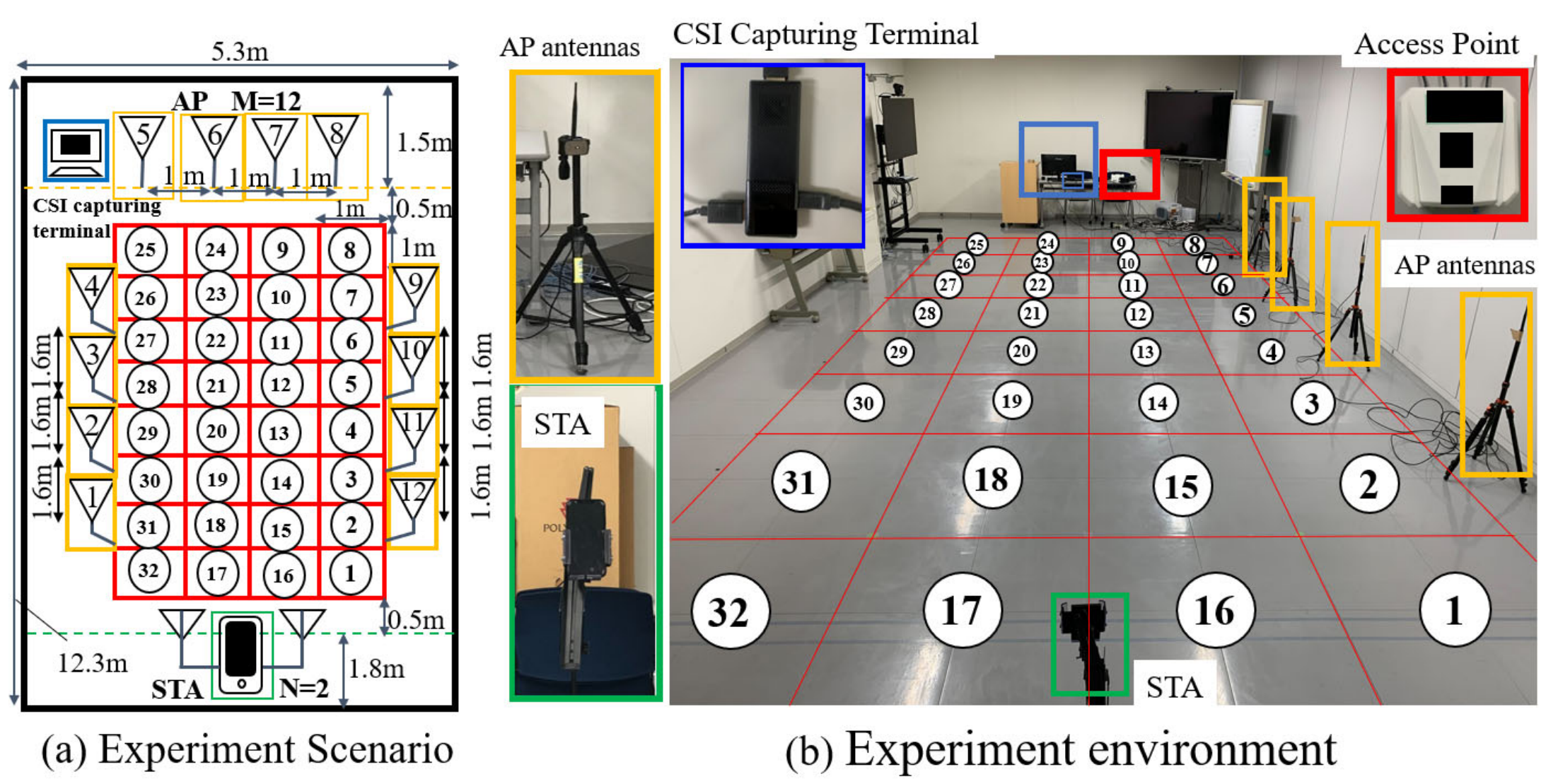} %
\caption{Experiment scenario and setup, where $M_a$ AP antenna positions are selected among $N=12$ candidate positions.}
\label{setup}
\end{center}
\end{figure}

To clarify the effectiveness of the proposed approach, experiment-based evaluations are conducted in an indoor environment. 
The block diagram of an IEEE802.11ac-based WLAN system is the same as that portrayed in Fig.~\ref{Block diagram}, where an IEEE802.11ac-based AP and STA (WLAN devices) are used~\footnote{
\red{The proposed method works well even in the presence of other IEEE802.11bgn APs, because the CSI measuring terminal can capture IEEE802.11ac-based packets correctly. In this experiment, the CSI packets were successfully captured, although various APs were detected in the experiment room.}
}. 
The experiment scenario and setup are portrayed respectively in Figs.~\ref{setup}(a) and \ref{setup}(b).
%
The AP has an antenna array. 
Each antenna is connected to the AP using coaxial extension cables. 
The AP antennas and the STA (iPhone XR) are placed on top of the tripods, as shown in Fig.~\ref{setup}.
Feedback frames containing the CSI (i.e., BRWs) are sent regularly from the STA. 
Wireless sensing functions, including our designed algorithms, are implemented on a stick-type computer (Compute Stick STK2M364CC; Intel Corp.) with an IEEE802.11ac wireless interface used as a CSI acquisition terminal as depicted in Fig.~\ref{Block diagram}. 
%
%
To build the overall CSI database of $M=12$ antennas for evaluating the achieved performance of all antenna placement patterns, CSI measurements are taken separately for three antenna subarrays, i.e., $(\#1,\#2,\#3,\#4)$, $(\#5, \#6, \#7, \#8)$, and $(\#9, \#10, \#11, \#12)$ in Fig.~\ref{setup}(a)\footnote{In Fig.~\ref{setup}(b), antenna subset $(\#5, \#6, \#7, \#8)$ is used.}. 
The overall CSI database for $M=12$ antennas is constructed by merging them. 
%
$\Xi=0$ is used unless stated otherwise. 
\red{
To build the trained random forest model, we used the scikit-learn library in Python. 
In our evaluations, default parameter settings in random forest library are used as typical parameters. 
The main parameter settings are as follows: the number of trees in the forest is 100, the number of features to consider when splitting is 1, the minimum number of samples in an internal node is 2, and the minimum number of samples in an external node is 1.
}

\begin{table}[t]
	\begin{center}
		\caption{Experiment setup}
		\label{jikken_syogen}
		\begin{tabular}{ | c | c |}  \hline\hline
		 {Number of antennas at AP} & {$M_a=4$} \\ \hline
		 {Number of antennas at STA} & {$N=2$} \\ \hline
		 {Antenna height [m]} &{ 1 } \\ \hline
		 {Antenna spacing [m]} &{ $d=1$ } \\ \hline
		 {Bandwidth [MHz]} & { 20 } \\ \hline
		 {Center frequency [GHz]} & { 5.18 } \\ \hline
		 {Number of subcarriers} & { 52 } \\ \hline
		 {Machine learning model} & {Random forest} \\ \hline\hline
		\end{tabular}
	\end{center} 
\end{table}

\begin{figure}[t] 
\begin{center}
\includegraphics[width = 8.5cm]{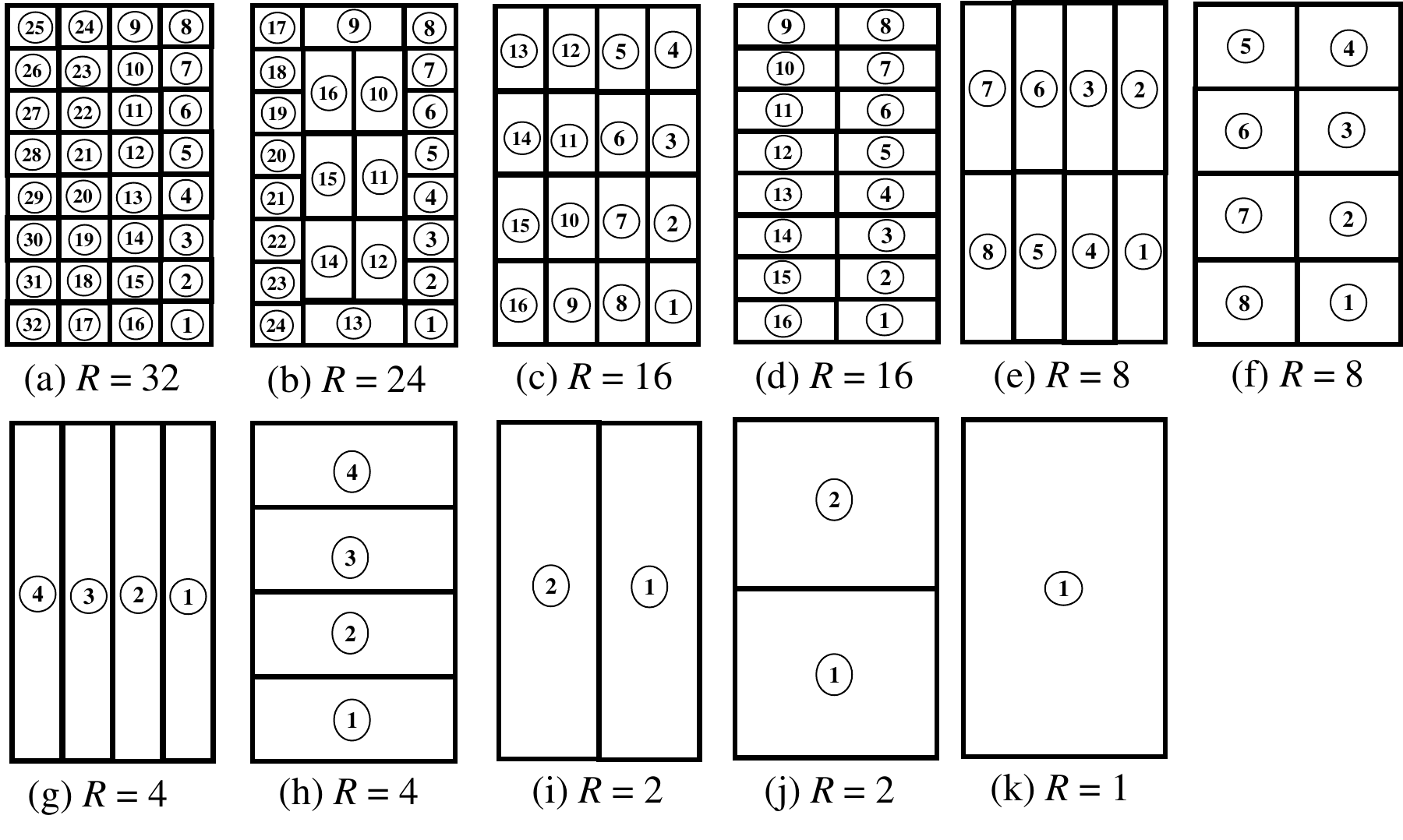} %
\caption{Area partition patterns, where $R$ denotes the number of partitioned areas. (a)--(k) correspond to different area partition patterns. }\label{region}
\end{center}
\end{figure}

To evaluate the relation between antenna positions and the achieved localization accuracy, four antenna elements are selected among 12 antennas. Then training and object detection are done using the corresponding CSI extracted from the database. 
Possible combinations of selected antenna patterns are $_{12}C_4=495$. Note that there is no need to consider all possible patterns in the actual situation, as the proposed algorithm can determine the appropriate antenna pattern without CSI in advance. 
Random forest is used as a supervised machine learning (ML) algorithm.
The extracted CSI is used for training the ML model and for conducting the object detection for selected antenna patterns. 
In training processing, the CSI acquisition terminal collects feedback frames that include CSI samples while the target moves within area-$r$. 
The acquired CSI samples are labeled as "$r$" and are used to train the corresponding ML models. 
The constructed dataset then trains the ML models for regression and classification, i.e., one classification model and $R$ area-wise regression models are constructed individually. 
The training and localization processes described above are done for all antenna combinations (495 patterns). 

This experiment employs a pre-processing scheme that was described for an earlier study~\cite{keiretsu}. 
Concretely, sequentially obtained $U$ CSI samples are concatenated as single feature information and are used for training the classification model and detecting the target, where $U$ stands for the number of CSIs to be concatenated, i.e., the length of the concatenated CSI.
Using the concatenated CSIs, the object state can be characterized in frequency and spatial domains, which is expected to improve the detection accuracy~\cite{keiretsu}. 
We designate this scheme as CSI concatenation, and $U=4$ is used.  


Figs.~\ref{region}(a)--(k) portray area partition patterns considered in this section, which correspond to the different numbers of partitioned areas: $R=1$, $2$, $4$, $8$, $16$, $24$, and $32$. 
After the classification model detects the area number at which the target person is located, the area-wise regression model corresponding to the detected area estimates the target's position. 
Actually, $R=1$ corresponds to the conventional regression localization scheme without classification ~\cite{regression2}. 

As a measure to evaluate the area detection accuracy where the target exists, 
the average detection probability $P_e$ is defined as the following conditional probability:
\begin{equation}
P_e=\frac{1}{R}\sum_{r=1}^{R} P_{r} = Prob (s_{out}={r}|s_{ans}={r}), 
\end{equation}
where $P_r$ denotes the detection probability at area $r$.  
$S_{out}$ and $S_{ans}={r}$ respectively represent the ML decision result and the correct one~\footnote{\red{
This paper considers a multi-area classification and regression problem and for single target detection. From the principle, the problem can be extended to multiple target detection at the cost of required complexity if the corresponding CSI dataset for multiple targets is available. Regarding this difficulty, a theoretical framework for a multiple target detection has been presented recently in the literature~\cite{Bartoletti}. Extension to multiple target detection is an item for future study.
}}.

In addition, the average error distance between the estimated target position and the correct one is 
\begin{equation}
    \bar{\epsilon}=E\left[\sqrt{(x-x_0)^2+(y-y_0)^2}\right],
\end{equation}
where $(x,y)$ and $(x_0,y_0)$ respectively represent the two-dimensional coordinates of the estimated target position and the correct one\footnote{\red{This paper has evaluated the localization performance in terms of error distance. The results in terms of error distance can be converted to those in terms of root mean square error (RMSE). 
We have confirmed that the similar CDFs can be obtained as a function of RMSE.}}. 
%
%

\begin{figure}[t] 
\begin{center}
\subfigure[$U=1$ (without CSI concatenation)]
{
\includegraphics[width = 7.0cm]{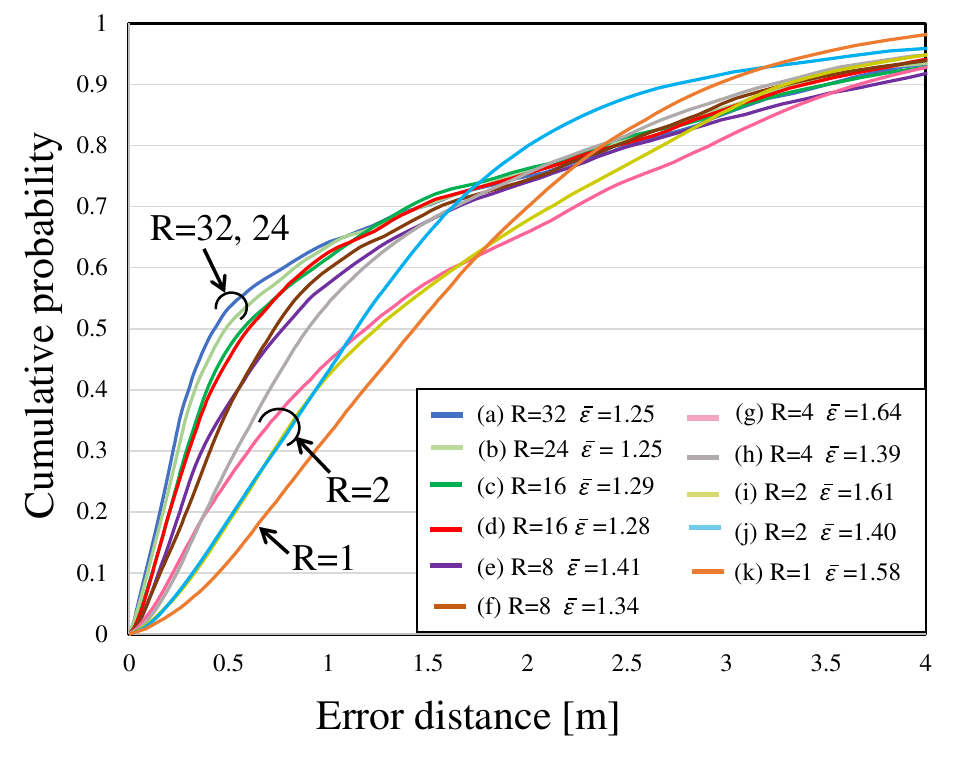}
\label{Fig7a}
}\\
\subfigure[$U=4$ (with CSI concatenation)]{
\includegraphics[width = 7.0cm]{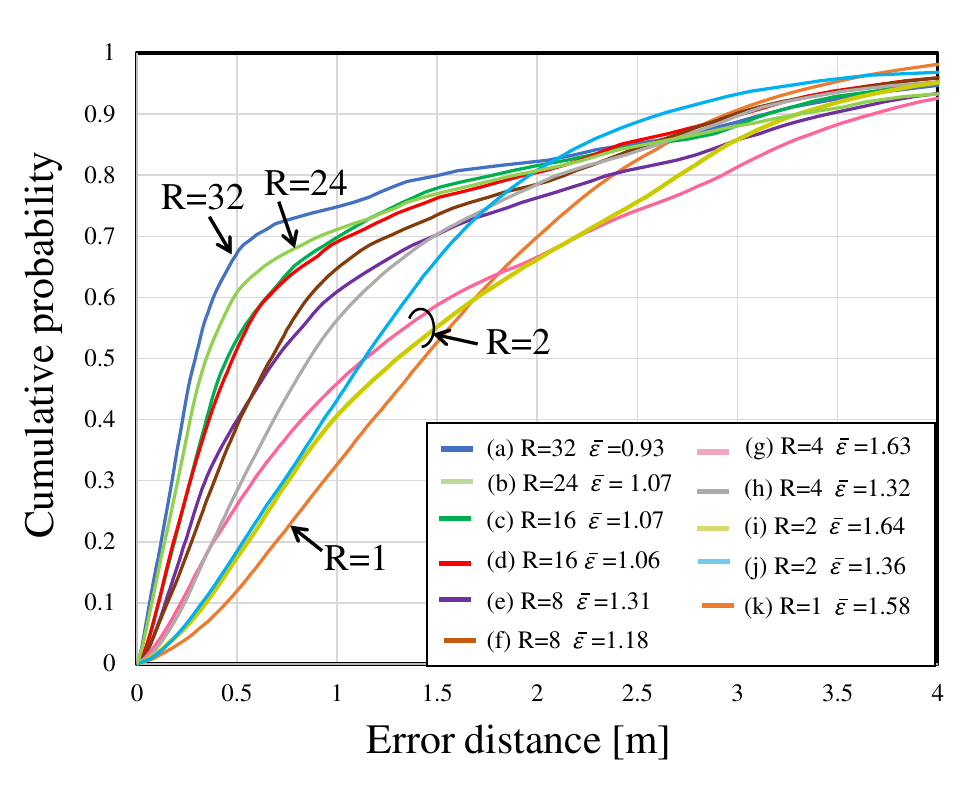}
\label{Fig7b}
}
\caption{Cumulative distribution of error distance in cases with spatial partition-wise regression: $R=1$, $2$, $4$, $8$, $16$, $24$, and $32$.}\label{Fig7}
\end{center}
\end{figure}

\begin{figure}[t] 
\begin{center}
\subfigure[$U=1$ (without CSI concatenation)]
{
\includegraphics[width = 7.0cm]{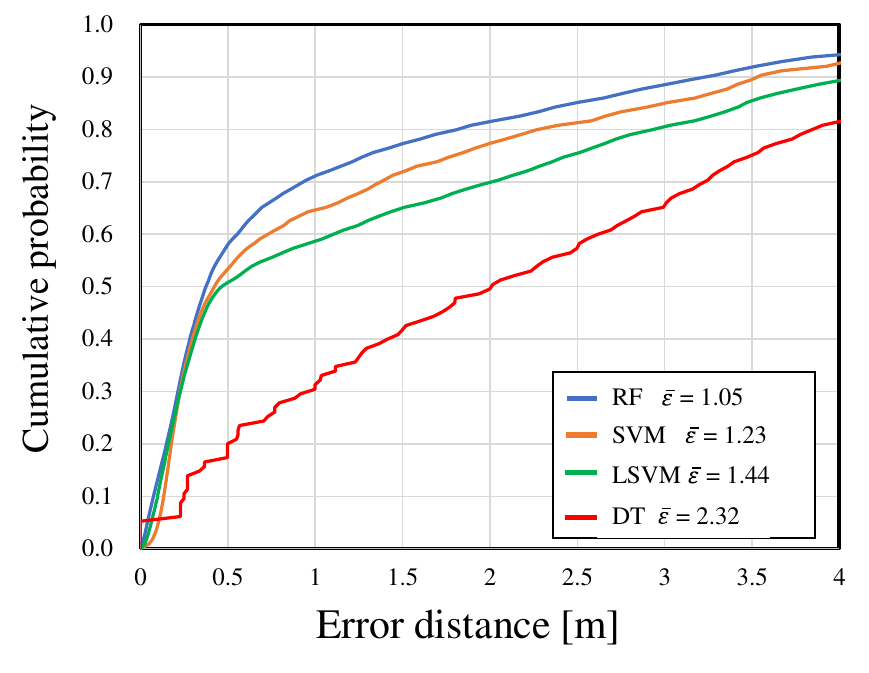}
\label{Fig8a}
}\\
\subfigure[$U=4$ (with CSI concatenation)]{
\includegraphics[width = 7.0cm]{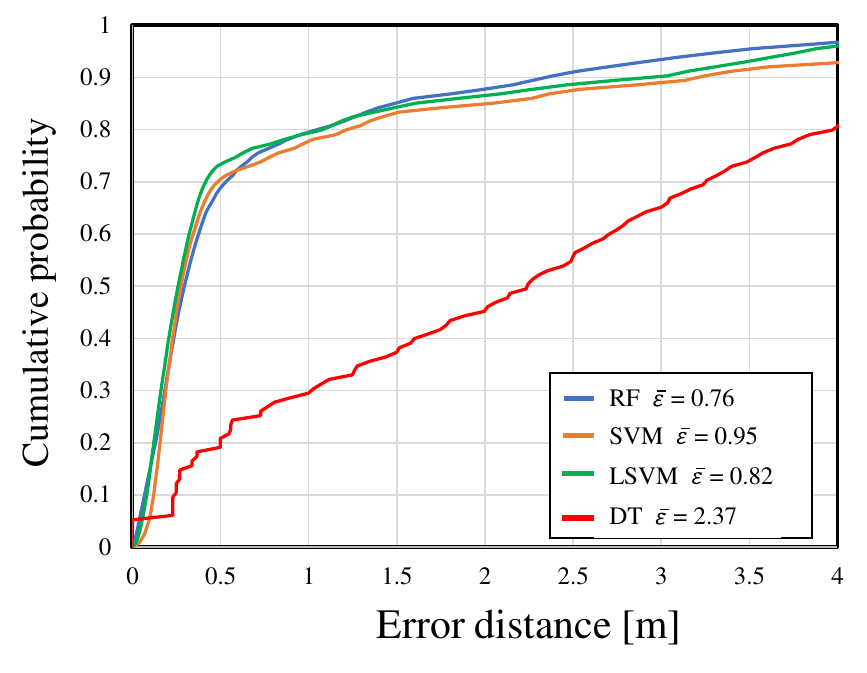}
\label{Fig8b}
}
\caption{Cumulative distribution comparison of error distance of the proposed scheme using other ML models in cases without ($U=1$) and without CSI concatenation ($U=4$), where the number of areas (labels) is $R=32$.}
\label{Fig8}
\end{center}
\end{figure}

\begin{figure}[t]  
\begin{center}
\subfigure[\red{Average detection probability}]
{
\includegraphics[width = 7.5cm]{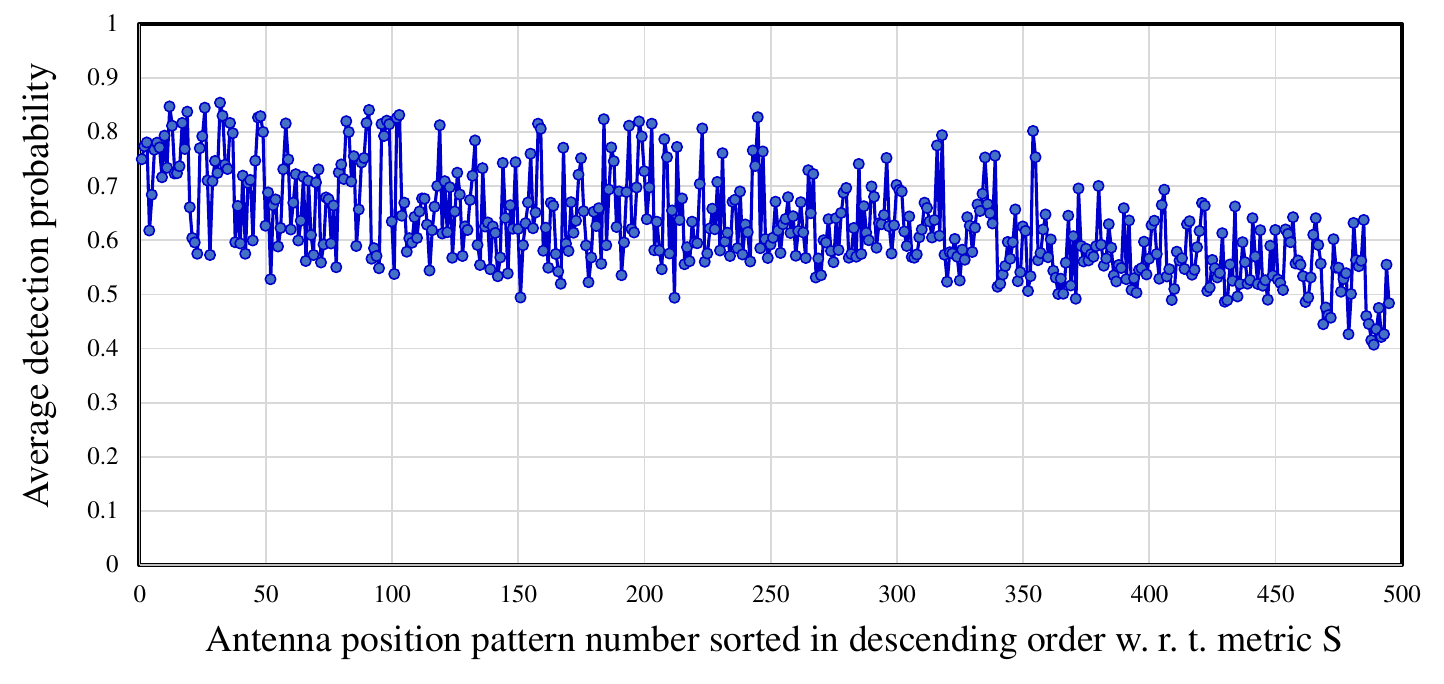}
\label{fig_svm1}
}\\
\subfigure[\red{Average error distance}]
{
\includegraphics[width = 7.5cm]{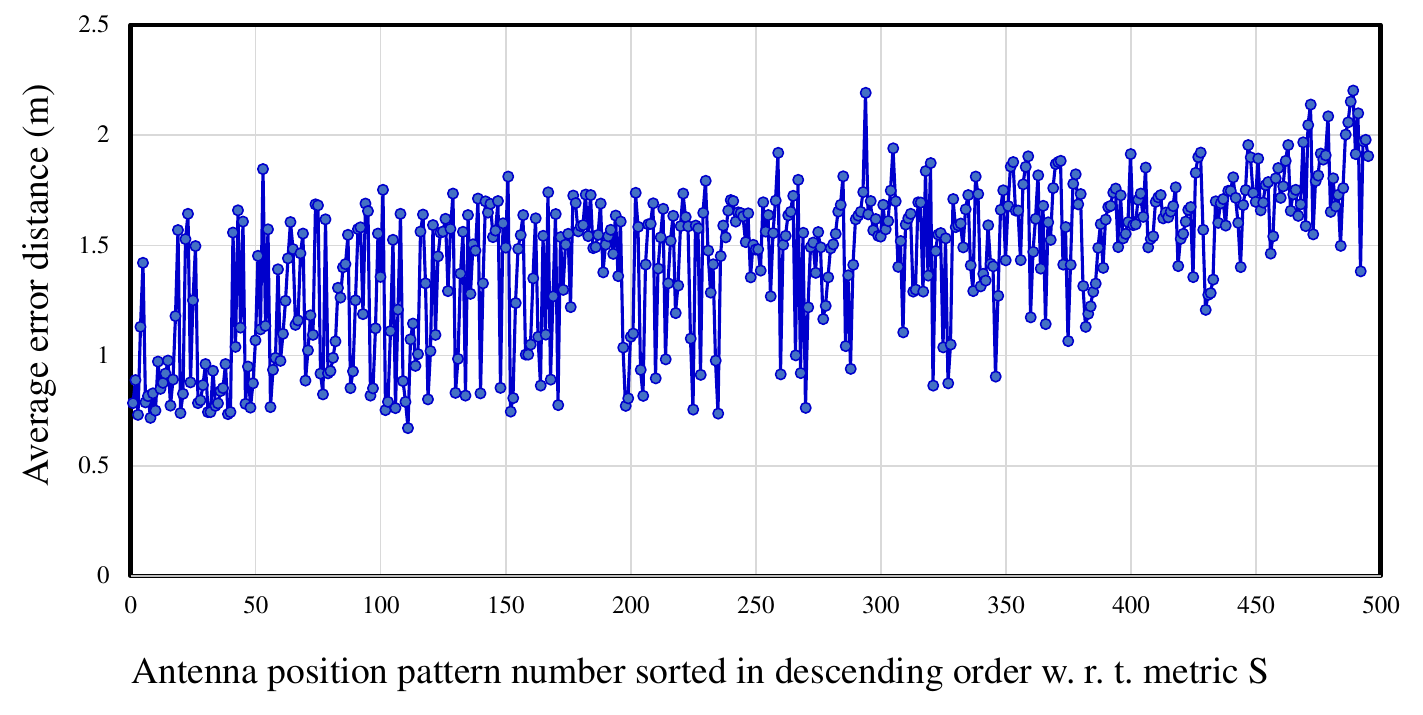}
\label{fig_svm2}
}
\caption{\red{Average detection probability and average error distance of the proposed scheme with SVM as a function of antenna position patterns sorted in descending order with respect to metric $S$.}}
\label{fig_svm}
\end{center}
\end{figure}

\begin{figure}[t] 
\begin{center}
\subfigure[Execution time for training both the regression model and classification model with $U=4$.]
{
\includegraphics[width = 7.0cm]{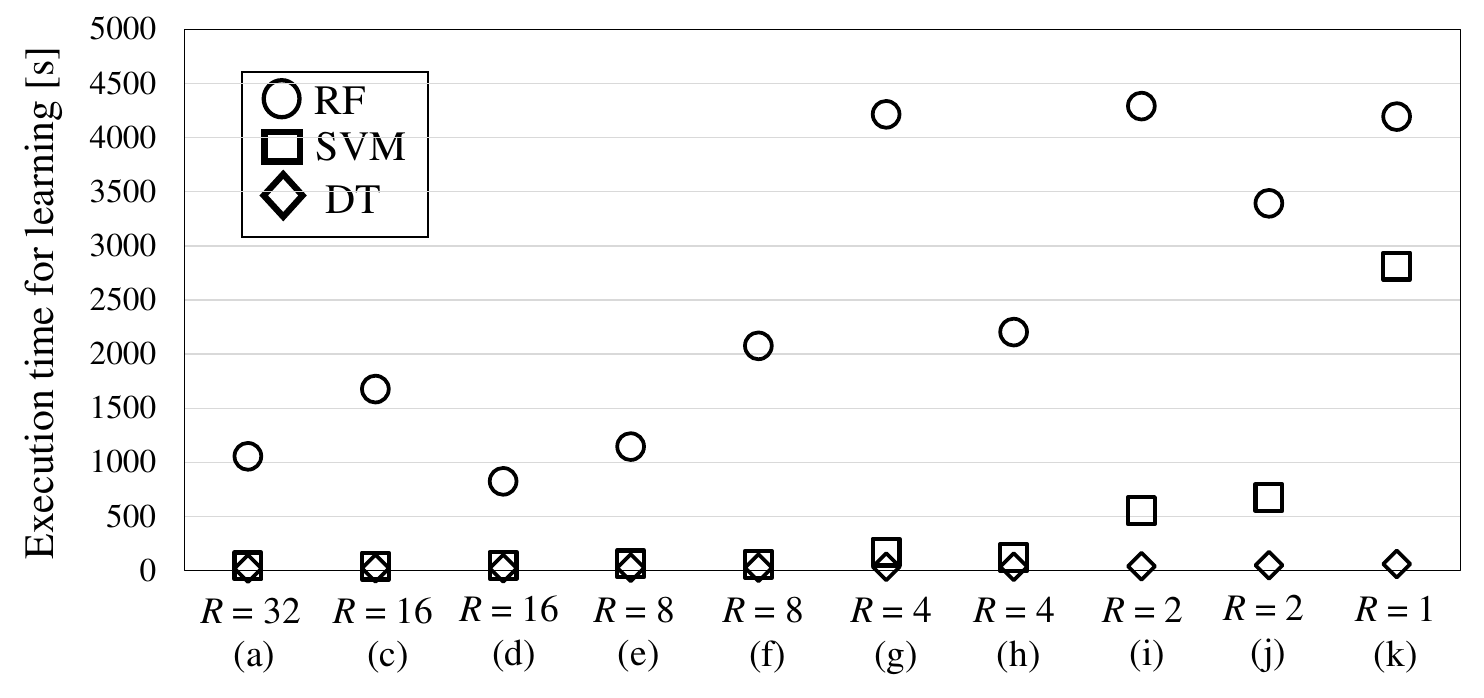}
\label{Fig9a}
}\\
\subfigure[Achieved average error distance as a function of execution time for testing, where $U=4$ and $R=32$]{
\includegraphics[width = 7.0cm]{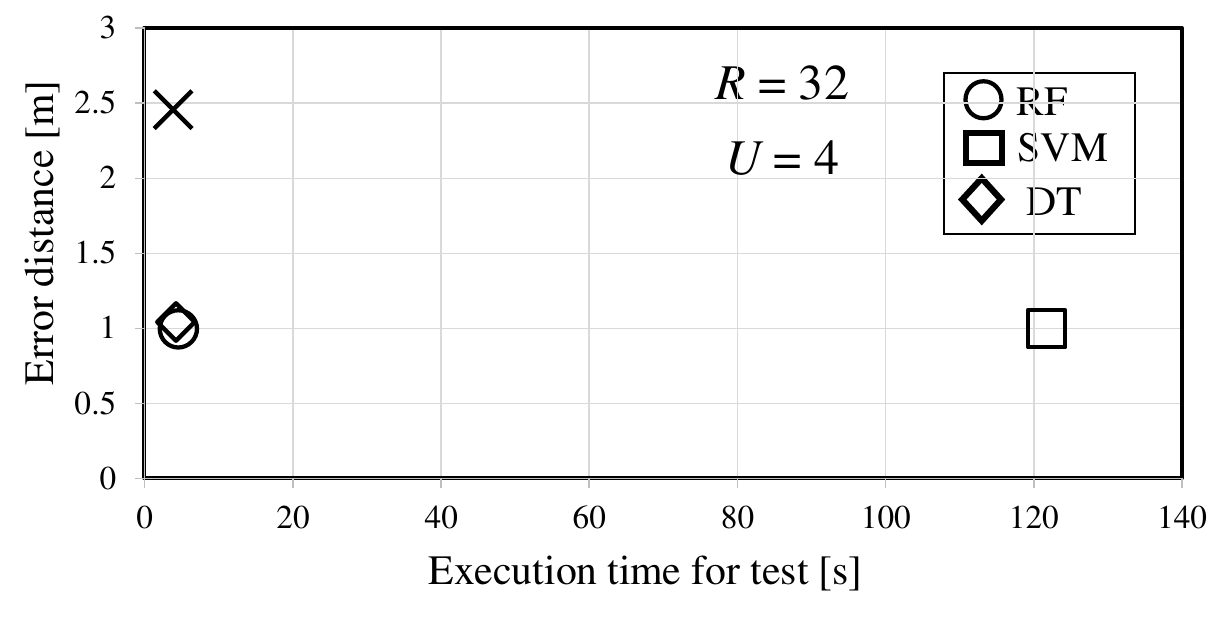}
\label{Fig9b}
}
\caption{Execution times for the regression model and classification model.}
\label{Fig9}
\end{center}
\end{figure}

\begin{figure}[t] 
\begin{center}
\includegraphics[width = 9cm]{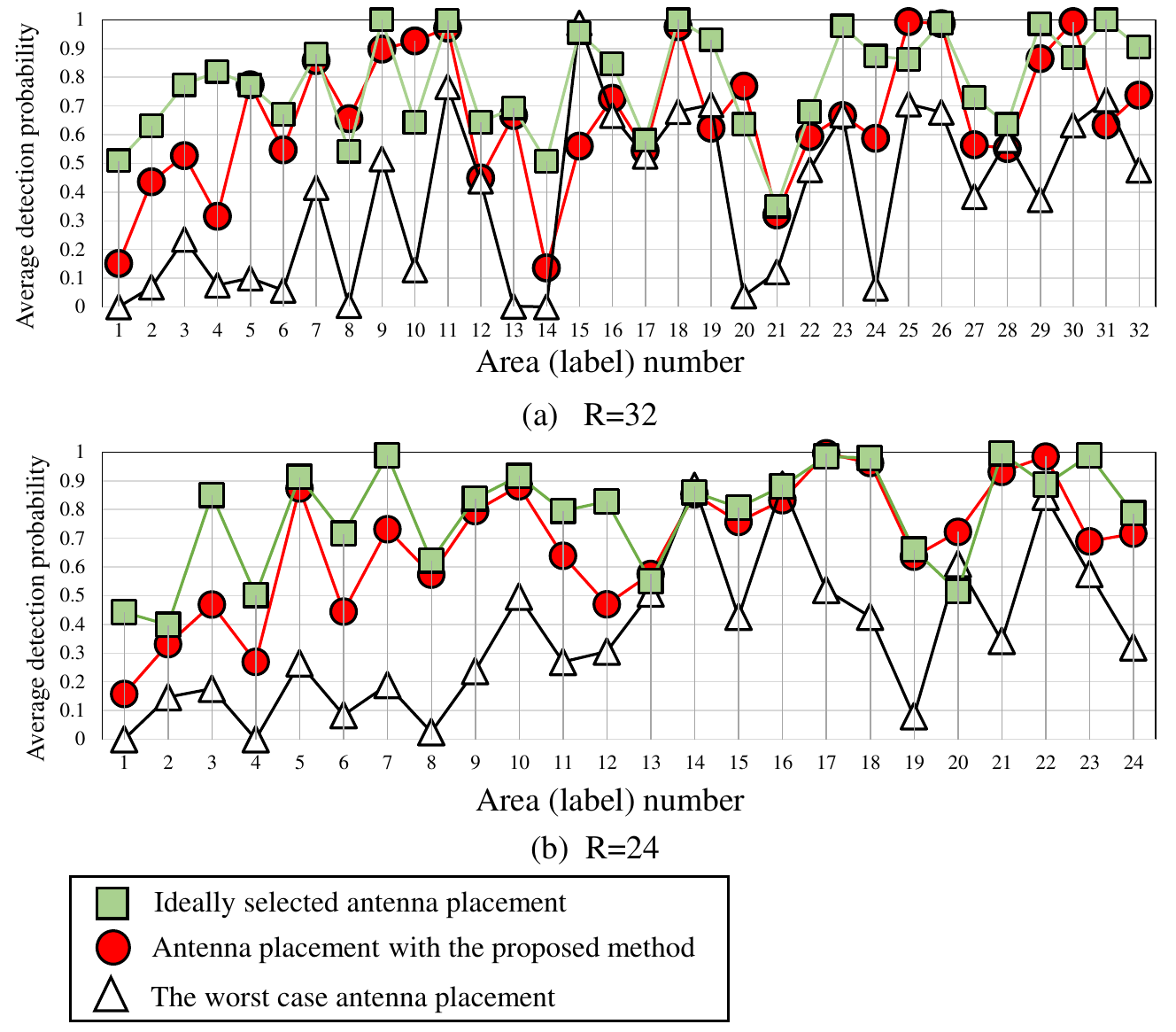} %
\caption{Area-wise average detection probability for $R=32$ and $24$, where Random forest classification with $U=4$ is used to detect the area number $r$. Here, $r=1,\cdots, R$, where $R$ denotes the number of areas.}
\label{Fig10}
\end{center}
\end{figure}

\subsection{Experimentally obtained results for localized regression}
To clarify the effect of the number of partitions $R$ on the achieved localization performance, the error distance performance is evaluated for a fixed antenna position using an antenna array pattern (\#5, \#6, \#7, and \#8). 
Fig.~\ref{Fig7} shows the cumulative distribution of error distance for the proposed scheme in cases $U=1$ and $4$ in terms of $R$ in the observation area, where $R=1$, $2$, $4$, $8$, $16$, $24$, and $32$ are used. 
Here, (a)--(k) in the legend correspond to those in Fig.~\ref{region}. 
$\bar{\epsilon}$ stands for the average error distance. 
In the smaller error distance region of Fig.~\ref{Fig7}(a), higher localization accuracy is obtained as $R$ increases. 
In contrast, in the higher error distance region, the
localization accuracy is degraded with the increase of $R$. 
This is true because, as $R$ increases, the probability of area detection errors increases. Consequently, it affects the localization accuracy by the regression model significantly.
However, in Fig.~\ref{Fig7}(b), the overall localization accuracy is improved even in the higher error distance region, in contrast to the case of $U=1$ because using the CSI concatenation ($U=4$) improves the area detection probability significantly.  
This finding implies that the proposed approach improves localization accuracy more effectively when using concatenated CSI as feature information. 

Figs.~\ref{Fig8}(a) and \ref{Fig8}(b) respectively show the cumulative distributions of error distance for various ML models in the case of the proposed scheme with $U=1$ and $4$. Here, $R=32$ is used. 
For comparison, we consider four typical ML models: Random Forest (RF), decision tree (DT)~\cite{RF&DT}, support vector machine (SVM), and linear SVM (LSVM). 
These figures show that the ML models, except for DT, achieve similar localization performance, i.e., approximately 0.5 m, with cumulative probability of 0.7 in case with $U=4$.  

\red{
To demonstrate the effectiveness of the proposed scheme with other machine learning models, we evaluate the average detection probability and average error distance of the proposed scheme when an SVM model is applied as another machine learning model applicable to both classification and regression problems. 
Figure~\ref{fig_svm} shows the average detection probability and average error distance of the proposed scheme with SVM as a function of antenna position patterns sorted in descending order with metric $S$. 
The result clarifies that the proposed scheme is able to successively select better antenna positions even when SVM is used as a classification and regression model. 
}

\red{Furthermore,} to examine the complexity of the proposed scheme, we evaluate the execution times necessary for training and localization processing. 
The execution time is the elapsed time measured for training and localization processing using the same workstation (Intel Core i7-10750H CPU and 64 GB memory). 
Fig.~\ref{Fig9a} shows the execution time for training regression and classification models with various values of $R$ in the case of $U=4$, where RF, SVM, and DT are used.  
Here, the dataset sizes are equal for all cases. 
It can be confirmed from this figure that the execution time is shortened as $R$ increases because the required complexity for training the regression models is reduced with the increase of $R$, whereas that for the classification model is almost identical, irrespective of $R$. 
Therefore, partitioning the observation area can reduce the complexity necessary for building a regression model, unlike a classification model. 
In other words, the observed channel characteristics in each area vary widely depending on the target's position when $R$ is set as a small value. Consequently, a more complex regression model must be built to represent it. 
Increasing $R$ tends to decrease the complexity to build the regression model. 

Fig.~\ref{Fig9b} presents the relation between the achieved error distance and the total execution time for the object localization, where $U=4$ and $R=32$ are used. 
The result demonstrates that the RF model achieves a good tradeoff between short error distance and the required time complexity~\footnote{
\red{Comparison of various typical ML models (RF, DT, LR, SVM, Linear SVM (LSVM), KN, and GNB) in terms of the average detection probability has been given in~\cite{keiretsu}. This paper employs the RF model trained with a small dataset as a reasonable model that achieves relatively higher localization accuracy and lower complexity for training and localization. }
}.

\begin{figure}[t] 
\begin{center}
\subfigure[$U=1$ (without CSI concatenation)]
{
\includegraphics[width = 8.5cm]{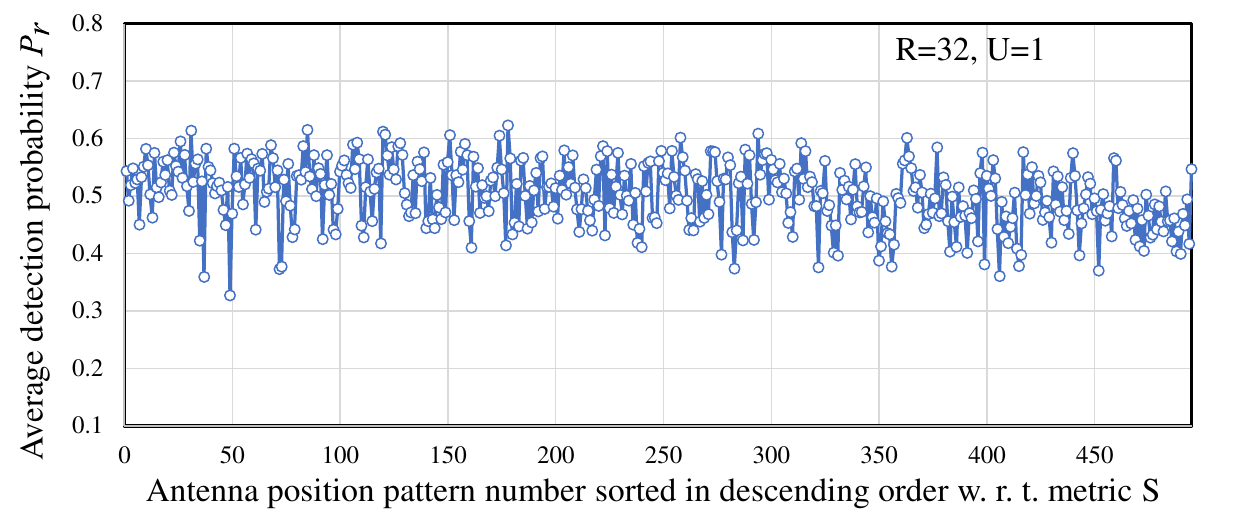}
\label{Fig11a}
}
\subfigure[$U=4$ (with CSI concatenation)]{
\includegraphics[width = 8.5cm]{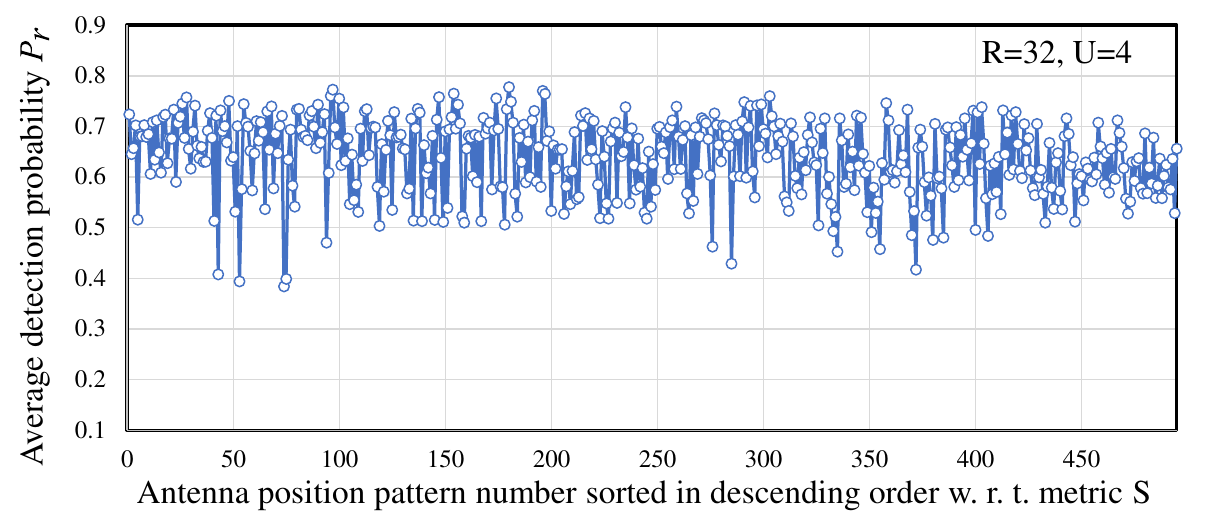}
\label{Fig11b}
}
\caption{Average detection probability for sorted antenna position patterns (sorted in descending order with respect to the evaluation metric $S$) in cases with $U=1$ and $4$, where $R=32$.}
\label{Fig11}
\end{center}
\end{figure}

\begin{figure}[t] 
\begin{center}
\subfigure[$U=1$ (without CSI concatenation)]
{
\includegraphics[width = 8.5cm]{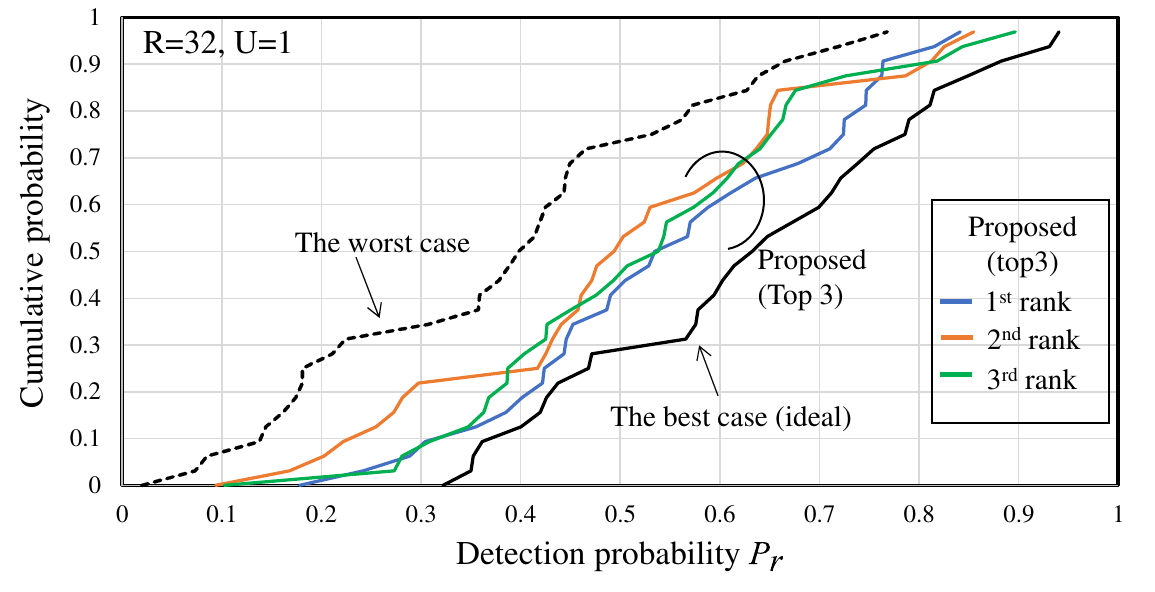}
\label{Fig12a}
}\\
\subfigure[$U=4$ (with CSI concatenation)]{
\includegraphics[width = 8.5cm]{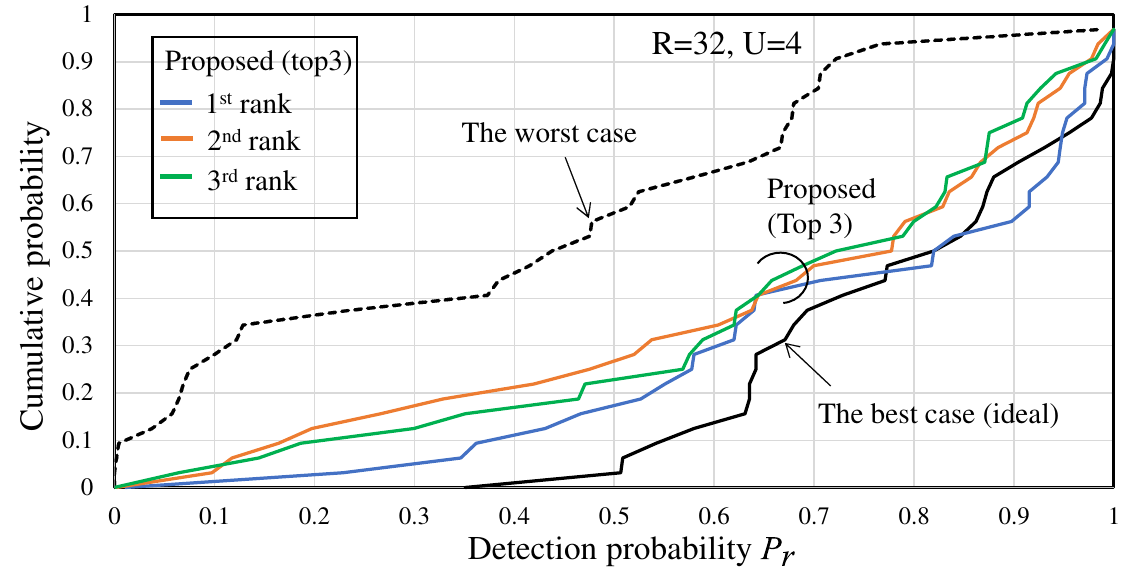}
\label{Fig12b}
}
\caption{Cumulative probability of area-wise detection probability for the proposed scheme in cases with $U=1$ and $4$, where $R=32$.}
\label{Fig12}
\end{center}
\end{figure}

\begin{figure}[t] 
\begin{center}
\includegraphics[width = 8.5cm]{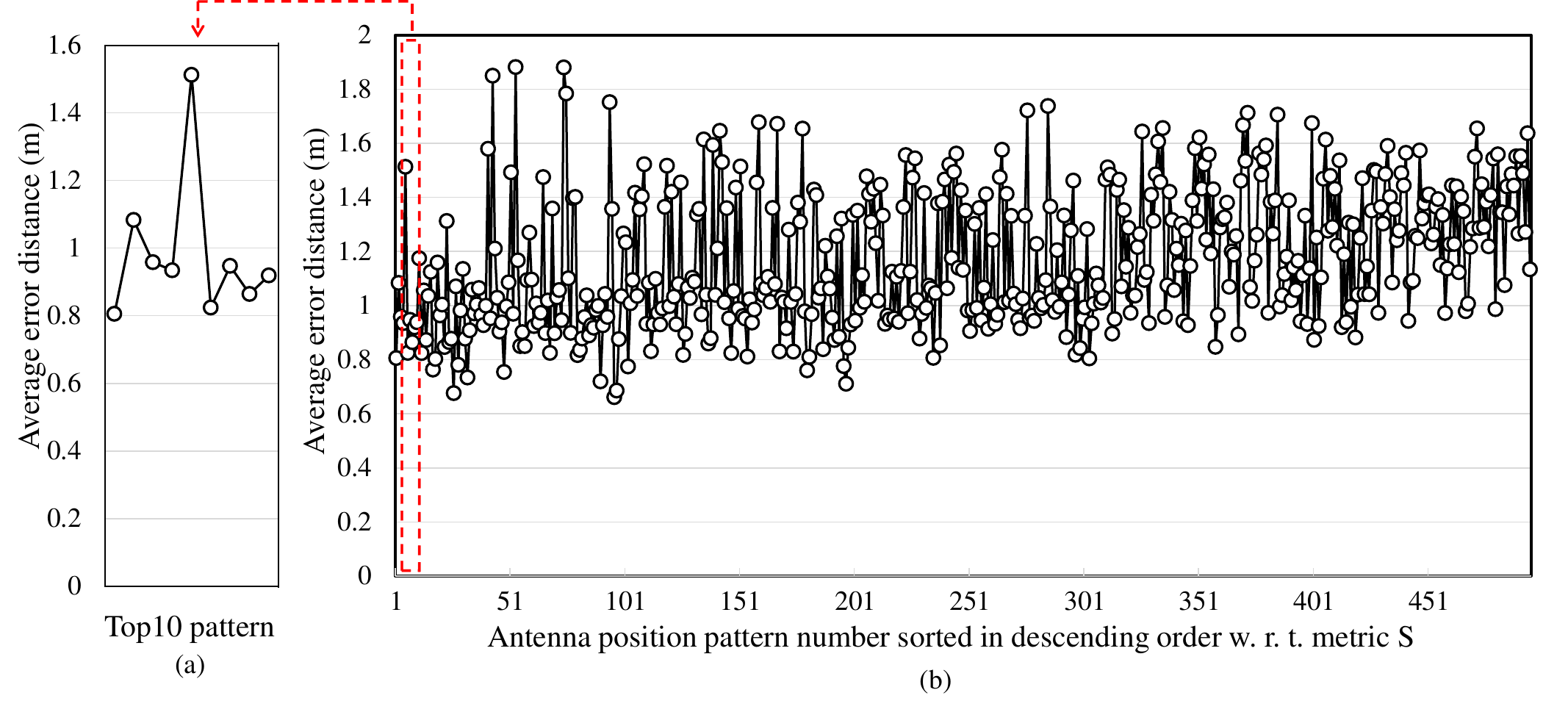} %
\caption{Average error distance for sorted antenna position patterns (sorted in descending order related to the evaluation metric $S$). The top 10 and all antenna patterns w.r.t. $S$ are presented in panels (a) and (b). }
\label{Fig13}
\end{center}
\end{figure}

\begin{figure}[t] 
\begin{center}
\subfigure[$U=1$ (without CSI concatenation)]
{
\includegraphics[width = 7.5cm]{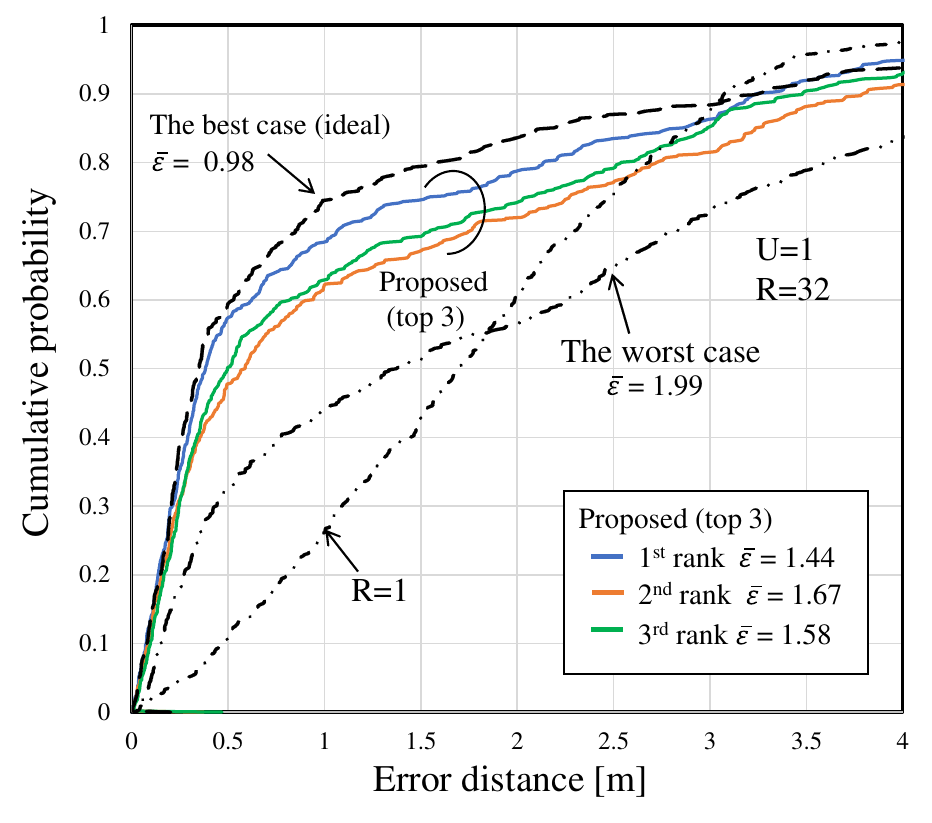}
\label{Fig14a}
}\\
\subfigure[$U=4$ (with CSI concatenation)]{
\includegraphics[width = 7.5cm]{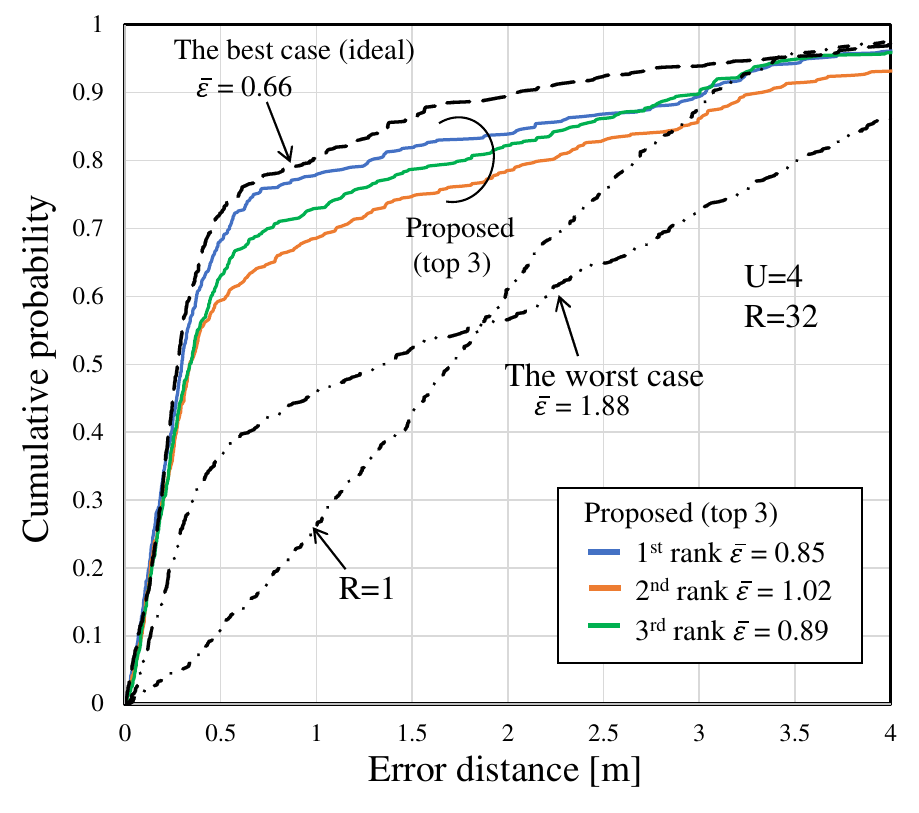}
\label{Fig14b}
}
\caption{Cumulative probability of error distance for the proposed scheme in cases with $U=1$ and $4$, where $R=32$.}
\label{Fig14}
\end{center}
\end{figure}

\begin{figure}[t] 
\begin{center}
\subfigure[$U=1$ (without CSI concatenation)]
{
\includegraphics[width = 7.5cm]{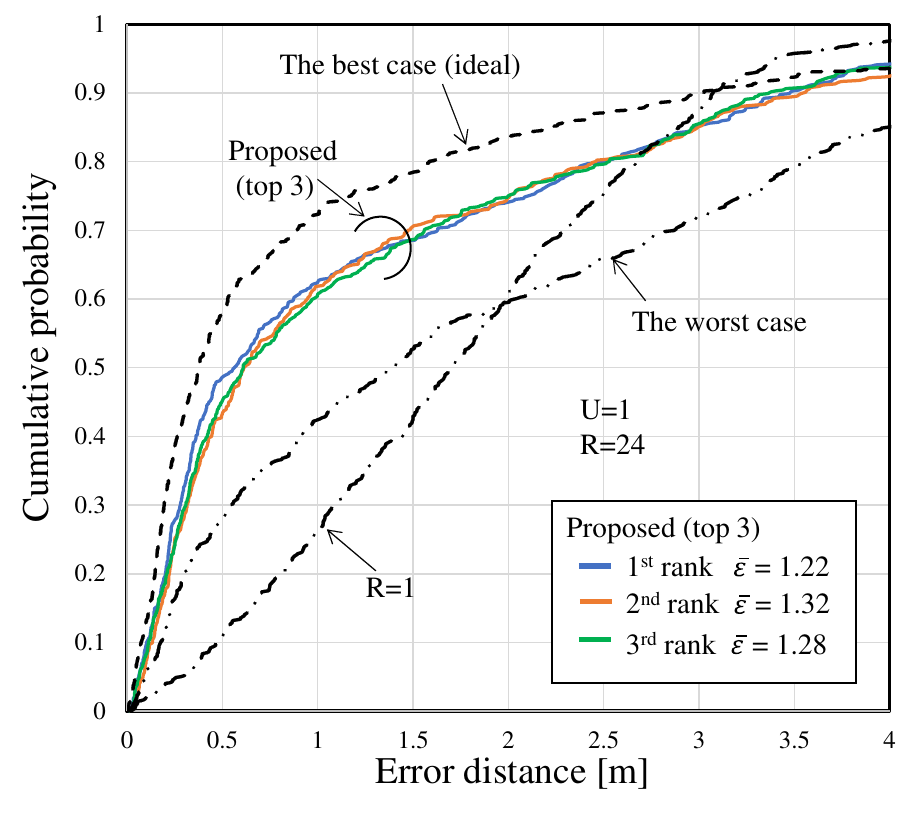}
\label{Fig15a}
}\\
\subfigure[$U=4$ (with CSI concatenation)]{
\includegraphics[width = 7.5cm]{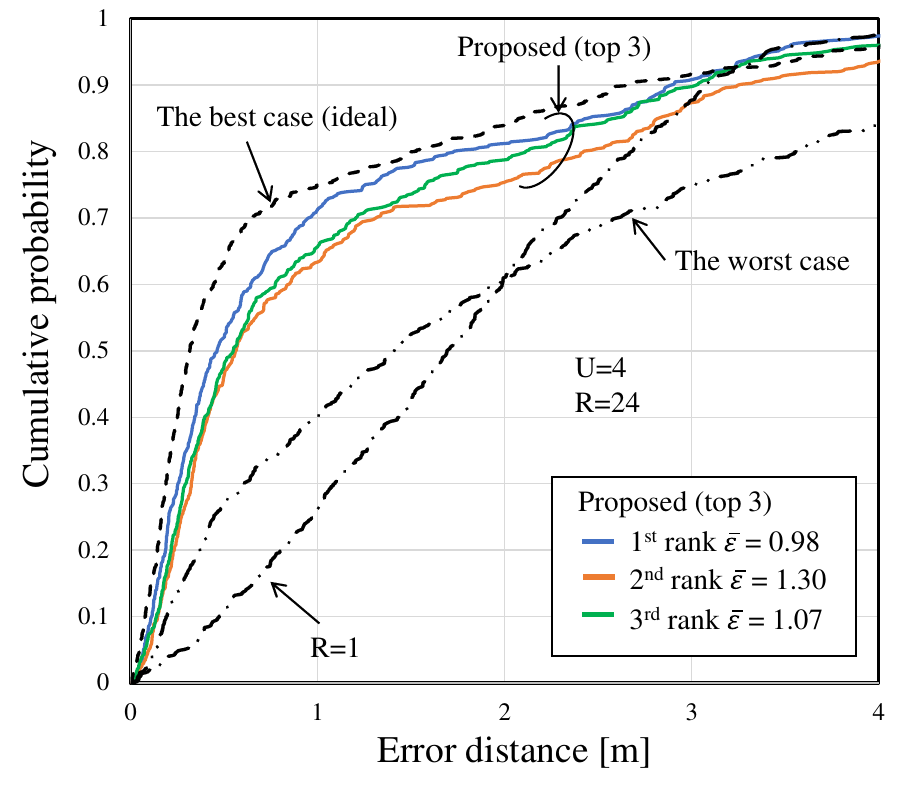}
\label{Fig15b}
}
\caption{Cumulative probability of error distance for the proposed scheme in cases with $U=1$ and $4$, where $R=24$.}
\label{Fig15}
\end{center}
\end{figure}

\begin{figure}[t] 
\begin{center}
\includegraphics[width = 7.5cm]{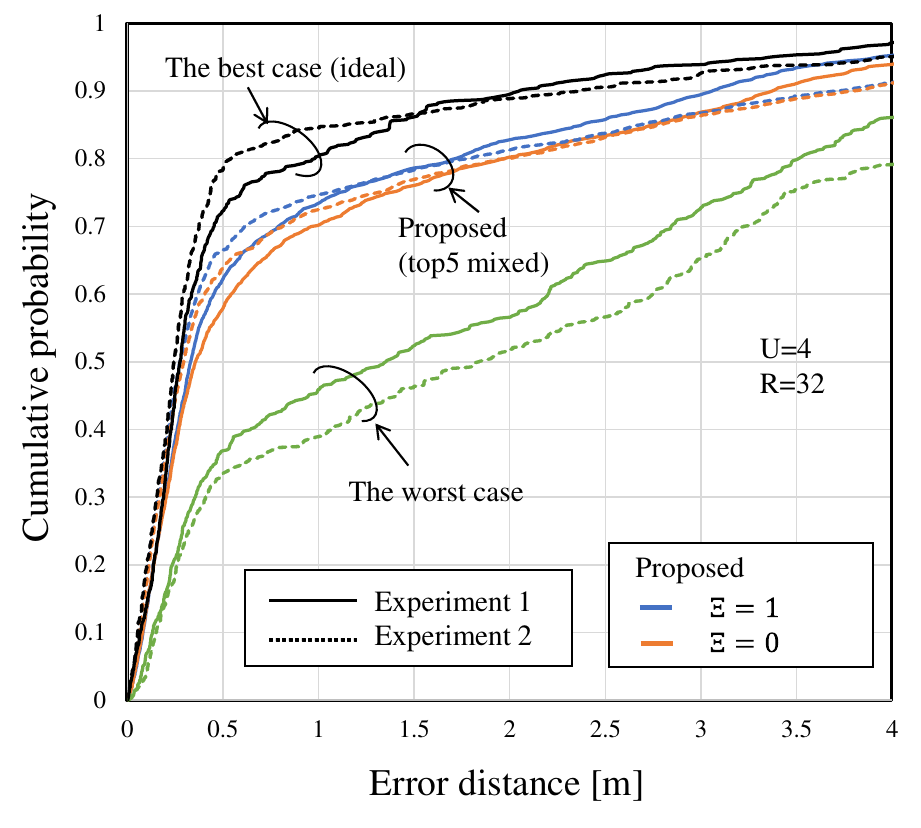}
\caption{Cumulative probability of error distance for the proposed scheme with $\xi=0$ and $1$, where $U=4$ and $R=32$.}
\label{Fig16}
\end{center}
\end{figure}

\subsection{Experimentally obtained results for the antenna placement algorithm}
Figs.~\ref{Fig10}(a) and \ref{Fig10}(b) respectively show the area-wise detection probability obtained when the AP antenna placement is determined with the proposed scheme in cases with $R=32$ and $R=24$. 
For comparison, the cases with the highest and lowest average detection probabilities are also shown respectively as results for the best and worst antenna patterns. 
The result indicates that the achieved detection probability is antenna-placement dependent. The proposed scheme achieves much better detection performance than the worst case. 

Figs.~\ref{Fig11}(a) and \ref{Fig11}(b) respectively present the average detection probability in terms of AP antenna placement patterns (sorted in descending order for the evaluation metric $S^{[b]}$) in cases with $U=1$ (without CSI concatenation) and $4$ (with CSI concatenation). 
Here, $R=32$ and $\Xi=0$ are used.
It is apparent from this figure that the proposed scheme achieves a higher detection probability using the antenna placement which achieves the best score. 

Fig.~\ref{Fig12} shows the cumulative distributions of area-wise detection probabilities for the proposed scheme in cases with $U=1$ and $4$, where $R=32$ is used.
The label "Proposed (Top 3)" denotes the results of antenna placement patterns with the top three evaluation scores (first-ranked to third-ranked results). 
The blue line shows the best result obtained using the proposed scheme  corresponding to the label "1st rank" in the legend. 
The cumulative probabilities for the best and worst cases are also shown for comparison. 
These results confirm that the proposed scheme achieves a higher detection probability when adequately determined antenna placement is used. 

Fig.~\ref{Fig13} shows the average error distance for antenna placement patterns sorted in descending order for evaluation metric $S$. Average error distances for the top 10 and all antenna patterns are shown respectively in Fig.~\ref{Fig13}(a) and~\ref{Fig13}(b). 
The antenna position is determined beforehand without measuring CSI. 
%
We can confirm that better localization accuracy is achieved when using the antenna position with the best metric value.  

Figs.~\ref{Fig14} and~\ref{Fig15} respectively present the cumulative distributions of error distance for the proposed scheme in the cases of $R=32$ and $24$. 
For comparison, the results obtained for the best and worst cases are also shown. 
The label "Proposed (Top 3)" denotes the results of antenna placement patterns with the top three evaluation scores similar to the previous figure. 
%
%
The cumulative distribution of the case with $R=1$ corresponds to the conventional regression-based method without classification in~\cite{regression2}. 
Here, panels (a) and (b) in the figures respectively show the results obtained using $U=1$ and $4$. 
The results verify that the proposed scheme achieves better error distance using the antenna placement pattern which minimizes the metric.  
 
To evaluate the effects of reflected waves on the performance of the proposed scheme, Fig.~\ref{Fig16} depicts plots of the mixed cumulative distribution of the error distance in the case of the proposed scheme with $\Xi=0$ and $1$ in cases using $U=4$ and $R=32$, 
for which results of the top five antenna placement patterns are mixed.
%
%
The attenuation factor is $r_1=1.0$. 
This figure clarifies that consideration of the influence of reflected waves (i.e., using $\Xi=1$) tends to achieve better localization accuracy than the case with $\Xi=0$ (i.e., considering only direct waves). 
To confirm the effectiveness of the proposed scheme, we conducted experiments (CSI measurements) twice on different days. These evaluated results respectively correspond to the solid lines (Experiment 1) and the dotted line (Experiment 2)\footnote{Experiment 1 uses the same CSI as results in the other figures, whereas Experiment 2 uses new CSI measurements.}.  
This result implies that the proposed scheme can determine antenna positions more effectively by tracing direct and reflected paths. 
Additionally, these findings confirm that similar trends are shown by the results of Experiment 1 and Experiment 2.

\begin{figure}[t] 
\begin{center}
\includegraphics[width = 8.5cm]{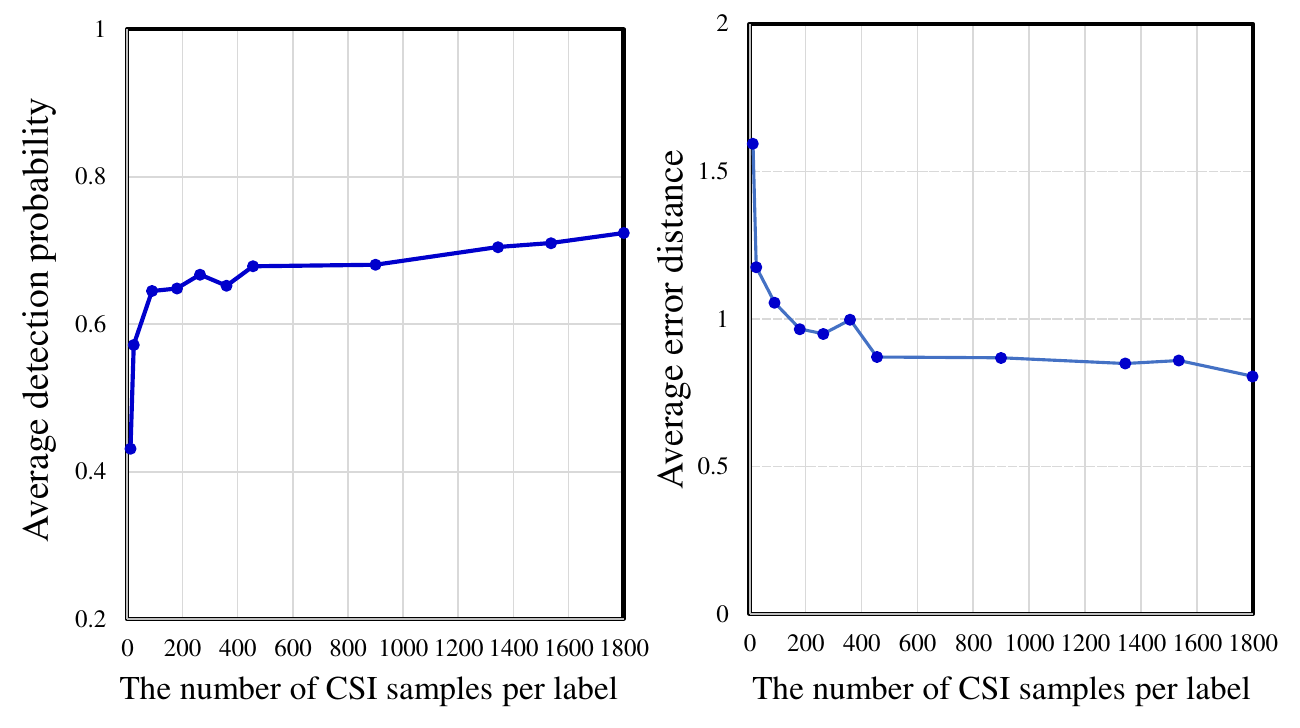} %
\caption{\red{Average detection probability and average error distance as a function of the number of CSI samplers per label, where $U=4$ is used.}}
\label{fig17}
\end{center}
\end{figure}

\begin{figure}[t] 
\begin{center}
\includegraphics[width = 8.5cm]{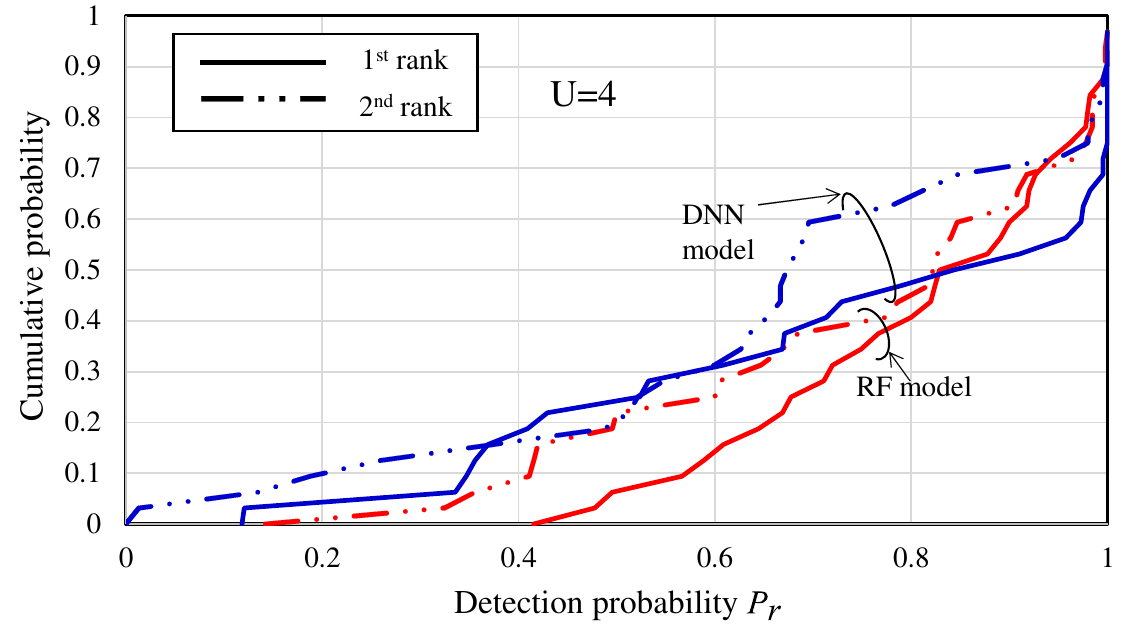} %
\caption{\red{Cumulative probability of area-wise detection probability for cases with the RF model and DNN model, where the concatenation size is U=4.}}
\label{fig_dnn}
\end{center}
\end{figure}

\begin{figure}[t]  
\begin{center}
\subfigure[\red{Experiment environment and setup}]
{
\includegraphics[width = 7.5cm]{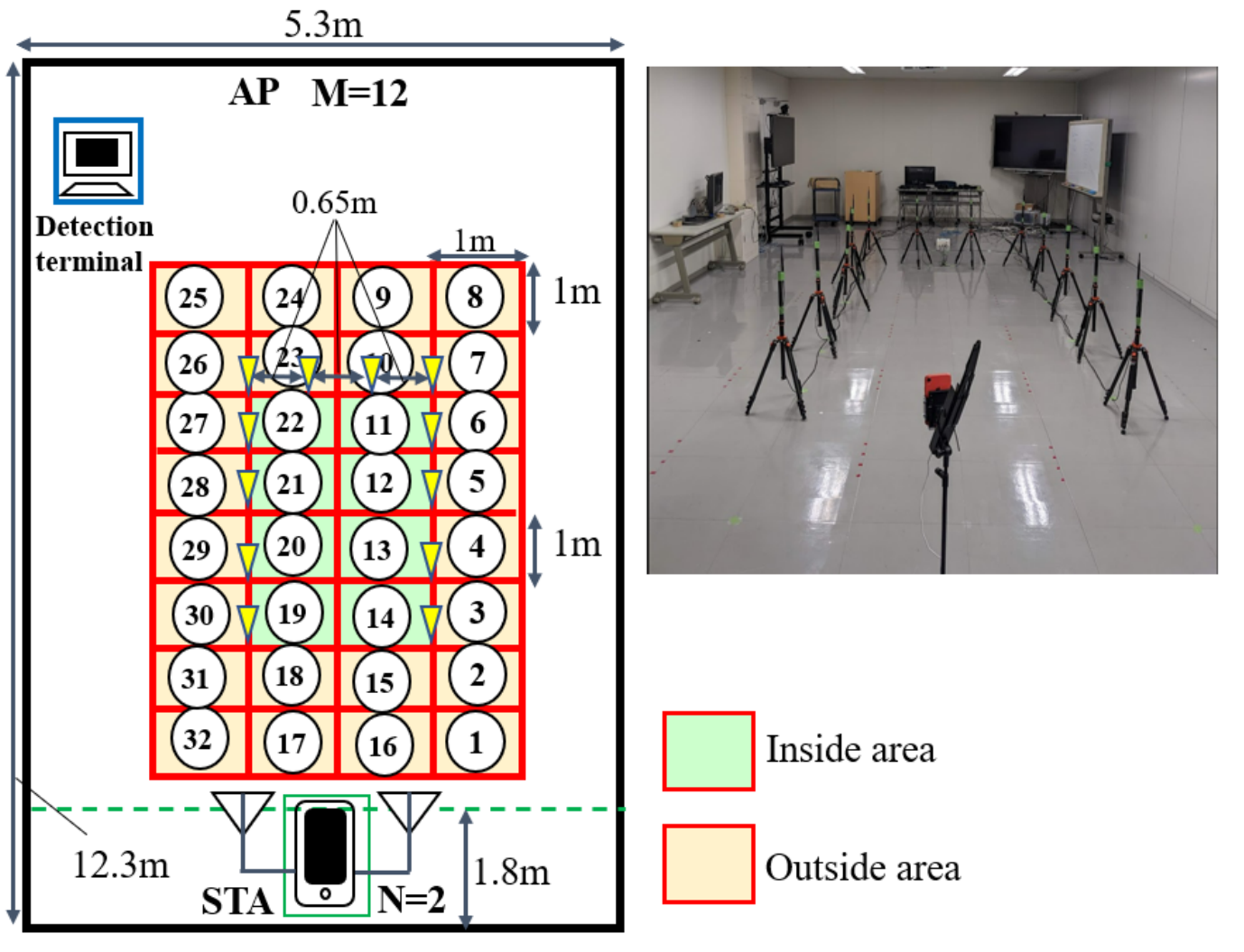}
\label{fig18a}
}\\
\subfigure[\red{Average detection probability as a function of antenna position patterns sorted by metric $S$}]
{
\includegraphics[width = 7.5cm]{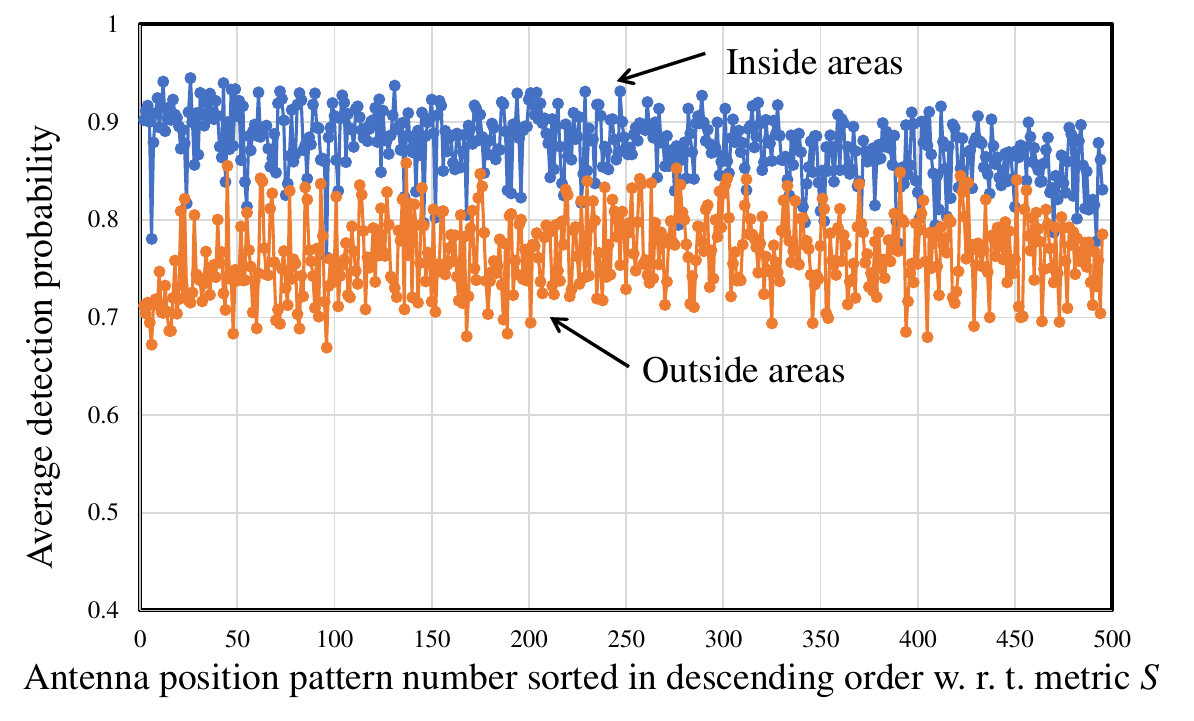}
\label{fig18b}
}
\caption{\red{Experiment scenario and setup where AP antennas are placed in the middle of the room.}}
\label{fig18}
\end{center}
\end{figure}

\begin{figure}[t]  
\begin{center}
\subfigure[\red{Experiment environment and setup}]
{
\includegraphics[width = 7.5cm]{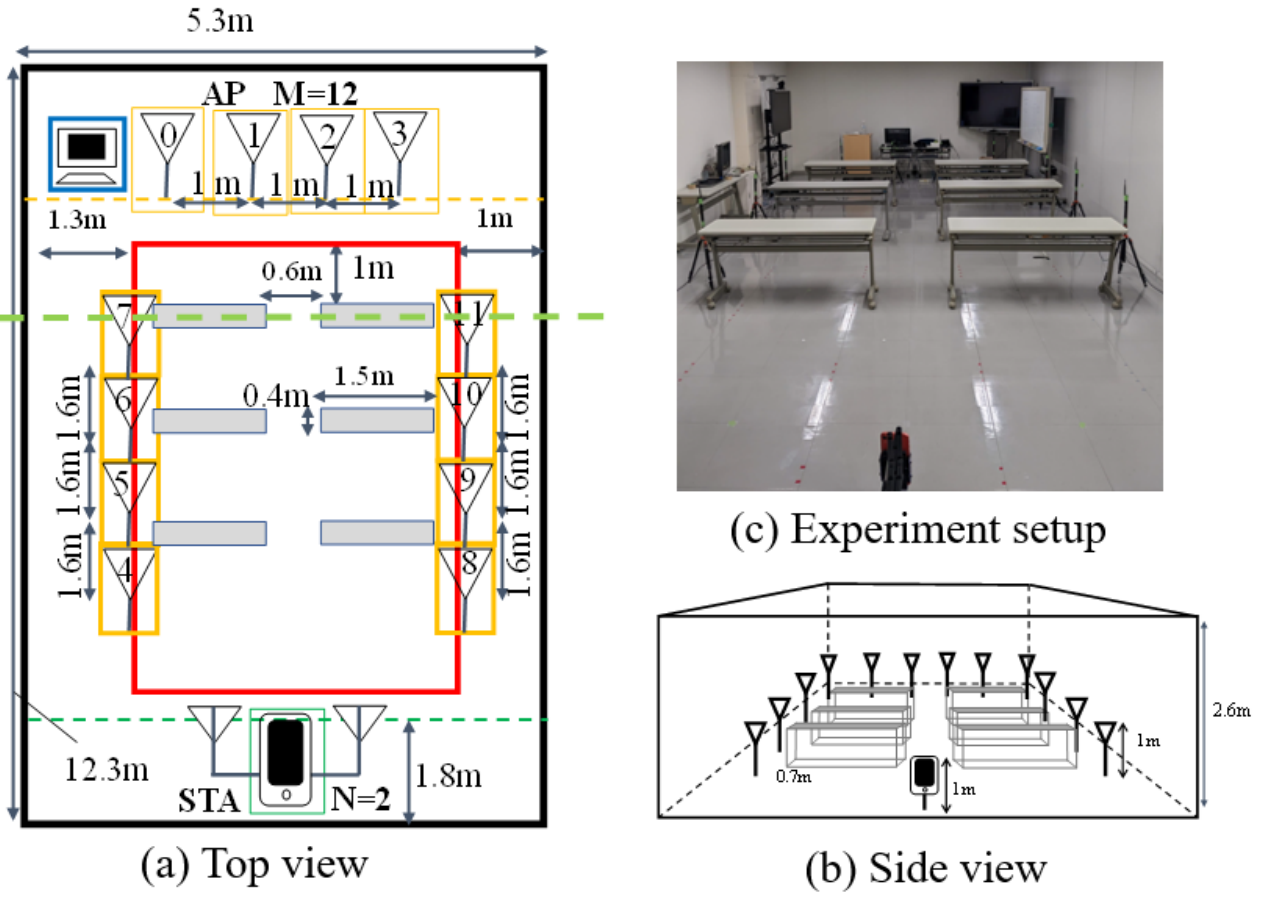}
\label{fig19a}
}\\
\subfigure[\red{Average detection probability as a function of antenna position patterns sorted by metric $S$}]
{
\includegraphics[width = 7.5cm]{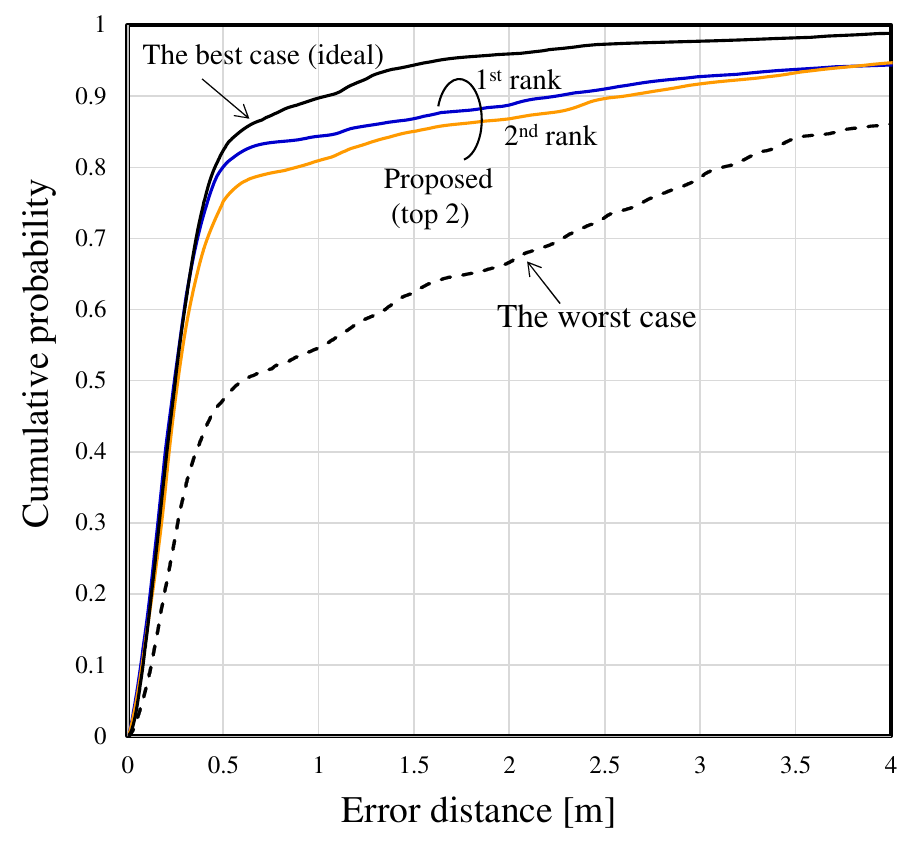}
\label{fig19b}
}
\caption{\red{Experiment setup and results in the scenario where there are tables in the room.}}
\label{fig19}
\end{center}
\end{figure}

\begin{figure}[t] 
\begin{center}
\includegraphics[width = 7cm]{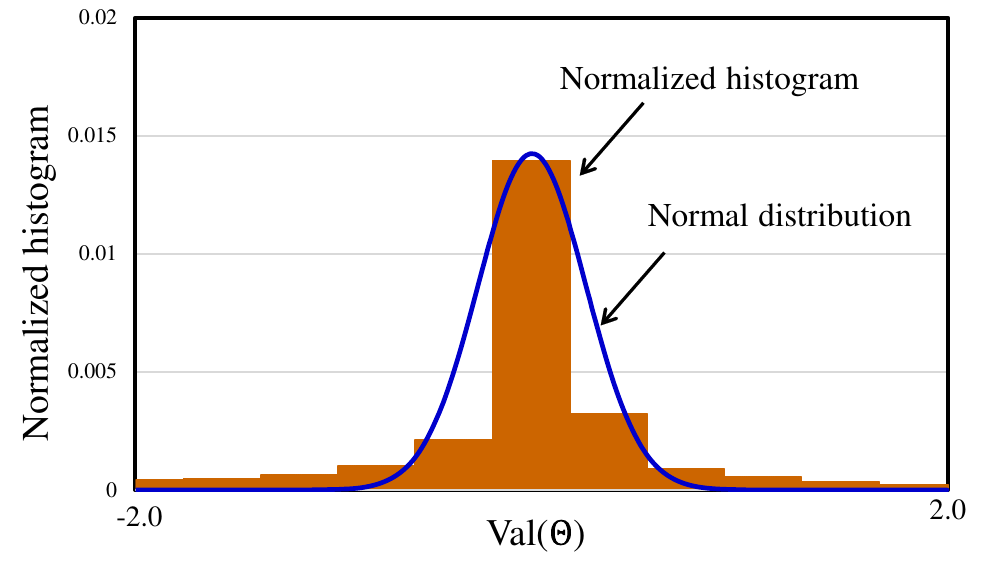} %
\caption{\red{Normalized histogram of Val($\Theta$) and fitted normal distribution.}}
\label{fig20}
\end{center}
\end{figure}

\red{
To clarify the impact of the number of CSI on localization performance, the relation between the number of training samples and achieved localization performance is evaluated. 
Figure~\ref{fig17} shows the average detection performance and the average error distance as a function of the number of training CSI samples per area in cases of U=4. 
These figures portray that the proposed scheme achieves similar performance even when the number of CSI samples per label is reduced significantly. 
The results reveal that the proposed method works well with a small dataset size.
}

\red{
The method in this paper aims at a lightweight approach with a small dataset and employed the RF model for this purpose. 
To compare the RF model with other models under the assumption that a small CSI dataset is used, we evaluate the average detection probability of a deep neural network (DNN) model trained on the same dataset and compare it with the RF model. Figure~\ref{fig_dnn} shows the cumulative probability of the area-wise detection probability for cases with the RF model and the DNN model, where the concatenation size is U=4. The labels "1st rank" and "2nd rank" denote the 1st and 2nd rank antenna patterns used in this paper. 
To solve the same multi-class classification problem, a DNN architecture consisting of two fully connected hidden layers and input and output layers is implemented as a supervised machine learning model (multi-class classifier)~\cite{Takahashi}. 
In this model, the number of components in the input layer, two hidden layers, and output layers are set to 3328, 128, 64, and 32, respectively. 
ReLU activation function is used. The RF model and the DNN model are trained on the same data set. 
It can be seen from this figure that both models achieve comparable localization performance when a small dataset is used. 
}

\subsection{\red{Experimentally obtained results for different scenarios}}

\red{
To discuss the required condition on antenna positions for the proposed scheme, we show experiment results when AP antennas are located in the middle area of the room. 
The experiment scenario is shown in Fig.~\ref{fig18a}. 
The average detection probability is calculated with the measured CSI dataset and plotted as a function of the antenna position pattern number sorted in descending order with respect to S. Figure~\ref{fig18b} shows the average detection probability results of the proposed scheme for inside area (green-colored areas: 11-14, 19-21) and outside area (yellow-colored areas: 1-10, 15-18, 23-32). 
The results in this figure indicate that the average detection probability for inside areas is improved by using the proposed antenna positions. 
On the other hand, there is no significant improvement in the average detection probability for outdoor areas by the proposed scheme. In addition, it is also clear that the average detection performance of inside areas is higher than that of outside areas for most antenna positions. 
Thus, we can conclude that the proposed approach works well when the observation areas are inside the AP antenna topology. 
In other words, the proposed scheme assumes that the AP antennas are located outside the observation area. 
}

\red{
To discuss the effectiveness of the proposed approach further, we show an experimental result in a room containing tables. The experiment scenario and setup are shown in Fig.~\ref{fig19a}. 
The experimental setup is the same as the other experiment scenarios in the paper except that there are tables in the room. 
Figure~\ref{fig19b} shows the cumulative probability of error distance for the proposed scheme in this scenario. 
It is confirmed from this result that the proposed scheme works well even when there are tables in the observation area of the room. 
These results imply that the proposed scheme is expected to work well in other indoor environments such as office rooms.
}

\subsection{\red{Statistical distribution of localization error in the proposed scheme}}

Let the $i$-th estimated two-dimensional coordinate and the correct one denote $(\hat{x}_i, \hat{y}_i)$ and $(x_i,y_i)$, respectively. 
Here, the localization errors in $x$-axis and $y$-axis are defined as $\epsilon_{x,i}=x_i-\hat{x}_i$ and $\epsilon_{y,i}=y_i-\hat{y}_i$, respectively. 
Hereafter, subscripts $x$ and $y$ are omitted for simplicity of notations, supposing that localization error $\epsilon_i$ is a normally distributed random variable with mean $\mu$ and variance $\sigma^2$. Based on this result, we will approximate $\epsilon_i$ as a normal distribution, and then define the following statistical parameter  
$\Theta=\sum_{i=1}^{N_{sample}}\frac{(\epsilon_i-\mu)^2}{N_{sample}}$.  
Here, we assume that the ensemble average of $\Theta$ approaches to $\sigma^2$. 
Let Val($\Theta$) denote the variance of $\Theta$. 
To analyze the statistical distribution of Val($\Theta$), the normalized histogram (empirical distribution) of Val($\Theta$) and fitted normal distribution are plotted in Fig.22, where Val($\Theta$) is calculated with area-by-area experimental results. Based on the statistical results in this figure, we can confirm that Val($\Theta$) behaves like a random variable. This fact suggests that the statistic of Val($\Theta$) can be discussed by empirically approximating it to a random variable following a normal distribution, e.g., Cramer Rao bound for a normally distributed random variable. 

\section{Conclusion\label{conclusion}}
As described herein, we have proposed an effective device-free WLAN-based localization scheme with a beam-tracing-based antenna placement optimization and localized regression, where concatenated BFWs are used as feature information for training ML-based models. 
In this scheme, the AP antenna element placement is found without knowing CSI by consideration of the relation between beam tracing and expected localization. 
Additionally, we have proposed the use of localized regression ML models to achieve more accurate localization with low complexity. 
We evaluated the effects of the proposed approaches by experimentation in terms of the object detection probability and error distance. 
The experimentally obtained results indicate that the proposed scheme is effective for appropriately selecting the antenna placement and consequently for detecting the target more accurately while reducing the required complexity compared to other reference schemes. 
Extensions of the proposed designs to various scenarios, such as wireless networks with joint communication and sensing with more powerful ML frameworks, are left as subjects of our future work.

\end{document}